\documentclass[preprint]{vgtc}               % preprint
\graphicspath{{figures/}{pictures/}{images/}{./}} % where to search for the images

\usepackage{times}                     % we use Times as the main font
\usepackage{tabu}                      % only used for the table example
\usepackage{booktabs}                  % only used for the table example
\usepackage{lipsum}                    % used to generate placeholder text
\usepackage{mwe}                       % used to generate placeholder figures
\usepackage{subcaption}

\usepackage{mathptmx}                  % use matching math font

\onlineid{1918}

\vgtccategory{Research}

\vgtcinsertpkg

\title{Through the Expert's Eyes: Exploring Asynchronous Expert Perspectives and Gaze Visualizations in XR}

\author{Clara Sayffaerth\thanks{e-mail: clara.sayffaerth@ifi.lmu.de}\\ %
        \scriptsize LMU Munich %
\and Annika Köhler\thanks{e-mail: annika.koehler@campus.lmu.de}\\ %
     \scriptsize LMU Munich %
\and Julian Rasch\thanks{e-mail: julian.rasch@ifi.lmu.de}\\ %
     \scriptsize LMU Munich %
\and Albrecht Schmidt\thanks{e-mail: albrecht.schmidt@ifi.lmu.de}\\ %
     \scriptsize LMU Munich %
\and Florian Müller\thanks{e-mail: florian.mueller@tu-darmstadt.de}\\ %
     \scriptsize TU Darmstadt}

\teaser{
  \centering
  \includegraphics[width=0.95\linewidth, alt={Two pictures of different perspectives on a self-build machine interface. The left picture shows the first-person perspective with two hands of a real person waiting to interact with the machine interface. Also, it shows two virtual floating hands of an instructor interacting with the machine interface, simulated by green-highlighted virtual replicas of the machine interface elements. In the right picture is a woman wearing a Head-Mounted Display standing beside the virtual expert dressed in blue overalls. The virtual expert is standing in front of the machine interface and demonstrates how to interact with the machine interface, highlighted again by green virtual replicas. The machine interface is mounted upright on the border of a table.}]{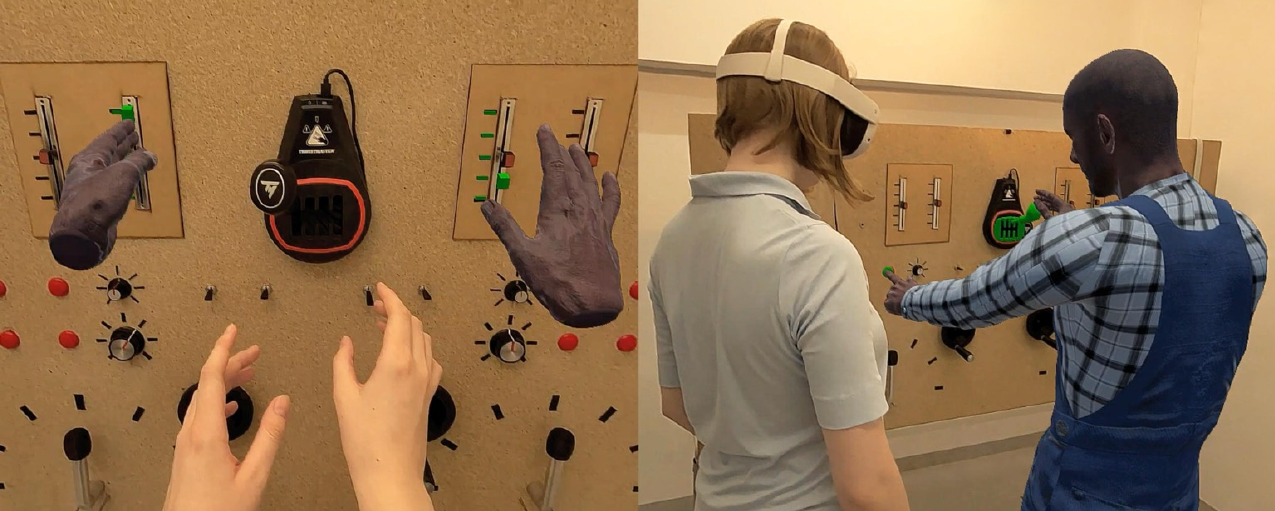}
  \caption{1st and 3rd person perspective of the expert during the machine task instructions in AR.
  }
  \label{fig:teaser}
}

\abstract{
  Transferring knowledge across generations is fundamental to human civilization, yet the challenge of passing on complex practical skills persists. Methods without a physically present instructor, such as videos, often fail to explain complex manual tasks, where spatial and social factors are critical. Technologies such as eXtended Reality and Artificial Intelligence hold the potential to retain expert knowledge and facilitate the creation of tailored, contextualized, and asynchronous explanations regardless of time and place. In contrast to videos, the learner's perspective can be different from the recorded perspective in XR. This paper investigates the impact of asynchronous first- and third-person perspectives and gaze visualizations on efficiency, feeling of embodiment, and connectedness during manual tasks. The empirical results of our study (N=36) show that the first-person perspective is better in quantitative measures and preferred by users. We identify best practices for presenting preserved knowledge and provide guidelines for designing future systems.
}


\begin{document}


\firstsection{Introduction}

\maketitle

%% \section{Introduction} %for journal use above \firstsection{..} instead
For thousands of years, the transfer of knowledge between generations has been a cornerstone of human civilization. The invention of writing, book printing, photography, film, and more recently the internet have fundamentally changed how knowledge is captured, recorded, handed on, and accessed. Despite this evolution of media, the loss of valuable information over time remains a major challenge to this day. Especially tacit knowledge, complex procedural and practical skills like two-handed tasks are challenging to explain without a present instructor~\cite{brunerFolkPedagogies1999} or through simple video streams~\cite{caoExploratoryStudyAugmented2020} that lack direct anchoring to the real world and guiding social cues.

As more people are leaving than joining the workforce, and instructing multiple people simultaneously is not effective~\cite{otsukiAssessmentInstructorCapacity2022}, there is an increasing need for asynchronous solutions that operate independently of a present expert. Future advancements in eXtended Reality (XR) and Artificial Intelligence (AI) systems promise to capture the expertise of specialists~\cite{chidambaramEditARDigitalTwin2022}, preserving their viewpoints and creating tailored, contextually anchored explanations for learners at any time~\cite{chidambaramProcessARAugmentedRealitybased2021}. Unlike current video-based solutions, such XR systems can decouple the perspective of the playback from the recording. This enables users to experience instructions from the experts' 1st person perspective (1PP) and various external 3rd person perspectives (3PP). This choice of perspective has practical relevance for learning success: Prior work indicated that for video tutorials, the 1PP performs significantly better for learning practical skills than the 3PP~\cite{fiorellaItAllMatter2017}. Further, by changing the perspective, the social connectedness~\cite{yongChangeSceneryTransformative2024} and the feeling of embodiment~\cite{liouDistinctionFirstpersonPerspective2023} can change, and therefore also the productivity and learning experience in general~\cite{banakouVirtuallyBeingEinstein2018}. Gaze is also an important factor that provides social cues and information about what someone is currently focusing on, improving the effectiveness, especially for novice learners~\cite{sungLearnersLearnMore2021}. However, this aspect gets lost in the 1PP, which is why researchers continued depicting the expert's head in asynchronous solutions~\cite{thanyaditXRLIVEEnhancingAsynchronous2022}, making it less realistic in a real environment. Other methods of visualizing the instructor's gaze have not been explored in this context. This raises the question of whether the improvement through 1PP also applies to XR and if additional social indicators enhance the effectiveness of asynchronous instructions.

In this paper, we explore effective methods for preserving valuable knowledge and ensuring its accurate, efficient, and enjoyable replay. Building on previous work, we examine the impact of different perspectives on learners and the potential benefits of gaze visualizations when following instructions to perform body-coordinated two-handed tasks. For this, we designed a study in which the participants saw and afterward imitated recordings of an expert. In the user study, we investigate how these factors influence learner efficiency, sense of embodiment, and social connectedness.

The contribution of this paper is twofold:

\begin{enumerate}
\item We contribute the results of a user study exploring the influence of perspective and gaze visualization on efficiency, sense of embodiment, and social connectedness for asynchronous XR instructions.
\item Based on the results, we provide guidelines for the future design of such systems.
\end{enumerate}

\section{Related Work}
Our work was based on already made approaches in the fields of knowledge transfer, synchronous and asynchronous XR, as well as different perspectives and gaze visualizations. In the following, we will further describe the research that influenced our decisions.

\subsection{Knowledge Transfer}
The way we learn has changed over the years through tools like color, writing, paper, and lately through sound and image recording devices. This transmission of facts or skills between entities is called knowledge transfer~\cite{røvikKnowledgeTransferTranslation2016}. When it comes to developing practical skills, spatial information and movement are important~\cite{bretzSelectionAppropriateCommunication1971}. However, Bandura's \textit{Social Learning Theory}~\cite{banduraSociallearningTheoryIdentificatory1969} describes how social factors can also play a role in knowledge transfer. Here, an observer learns through a model's actions and their consequences. Therefore, demonstrations of an expert are an important first step to achieving a skill, followed by imitation, practice, and promoting effective and active learning~\cite{brunerFolkPedagogies1999}. Moreover, social connectedness, as the feeling of belonging and closeness to others, can enhance the collaborative learning experience~\cite{fengEffectSocialCloseness2023}. When we go one step further, taking the perspective of an expert can not only increase our empathy for that person~\cite{yongChangeSceneryTransformative2024} but also positively influence the efficiency through the embodiment~\cite{banakouVirtuallyBeingEinstein2018}. These effects also occur when sharing gaze cues with each other~\cite{sungLearnersLearnMore2021}, which is helpful during two-handed tasks where other social cues such as hand gestures are difficult to perform. In addition, the style of an instruction can impact its success, for example, polite and less direct instructions are especially effective for learners with low prior knowledge or error-proneness~\cite{mclarenPoliteWebbasedIntelligent2011}. So not only does the plain understanding of the other person's actions affect our learning, but also the feeling of social connectedness during the process. Three-dimensional and dynamic XR seems like the most fitting approach to demonstrate manual tasks. Still, we need a thorough understanding of the social factors that can influence its effectiveness.

\subsection{XR and Time}
As we want to investigate which factors could influence the learning experience of XR instructions, we will further look into previous synchronous and asynchronous approaches for knowledge transfer.

\textbf{Synchronous XR} is used to gather people from different geographical locations at the same time. According to Schäfer et al.~\cite{schäferSurveySynchronousAugmented2022}, these applications can be categorized (among others) as \textit{Meeting}, \textit{Design}, and \textit{Remote Expert}. Over the next few years, the last category will become increasingly important as an aging population and growing technological advances mean that skilled workers will not always be available on-site.

In the field of telepresence, a considerable amount of research is being done to create the feeling for users of being in the same place as an expert. This involves not only making the exchange as close to reality as possible through the blending of multiple spaces~\cite{irlittiVolumetricMixedReality2023}, robotic extension~\cite{praveenaDesigningRoboticCamera2023}, tangible~\cite{villanuevaColabARToolkitRemote2022} and virtual objects~\cite{odaVirtualReplicasRemote2015}, but also preserving the most important aspects of their instructions by enhancing them with notifications~\cite{cidotaWorkspaceAwarenessCollaborative2016}, annotations~\cite{fakourfarStabilizedAnnotationsMobile2016} or cues~\cite{guntherExploringAudioVisual2018}. While these approaches are valuable, they reach their limits when faced with excessive demand, as in this scenario, an expert can act independently of location but not of time. As a result, the associated workload can lead to the expert being overburdened~\cite{otsukiAssessmentInstructorCapacity2022}. In addition, unique knowledge potentially gets lost with the expert. For this reason, researchers and industry also explore sustainable and time-independent solutions.

\textbf{Asynchronous} applications make information available anytime by preserving data. Comments~\cite{kimWinderLinkingSpeech2021}, recorded actions~\cite{chidambaramEditARDigitalTwin2022}, and tutorials~\cite{mayerImmersiveHandInstructions2023} allow for reviewing explanations and make knowledge more accessible through distribution and exchange. Thereby, asynchronous solutions can even outperform synchronous ones in these settings~\cite{thanyaditEfficientInformationSharing2018}, as confusing information can be avoided. However, no follow-up questions can be asked if the expert is no longer available. Especially in this case, asynchronous XR implementations have gained attention in the craft and technical context in recent years, as they can alter reality and, due to their three-dimensional nature, perform better in explaining practical knowledge compared to manuals~\cite{mayerImmersiveHandInstructions2023} or video tutorials~\cite{thanyaditXRLIVEEnhancingAsynchronous2022}. For this reason, research is being conducted to improve the XR knowledge transfer by adding wearables operating as sensors to capture the interaction with physical objects and as additional feedback~\cite{liuInstruMentARAutoGenerationAugmented2023}. Diverse playback options~\cite{chidambaramProcessARAugmentedRealitybased2021} optimize the arrangement of the individual task, minimize the chance of forgetting steps, and can affect co-presence in combination with the visualization of the instructor~\cite{thanyaditXRLIVEEnhancingAsynchronous2022}. Different notifications~\cite{marquesHowNotifyTeam2023}, annotations, and XR technologies~\cite{choAsynchronousHybridCross2023} help to focus and give additional information. Still, research on the body-related and social aspects of instructions that are crucial for practical learning and how they can be transferred and visualized for asynchronous XR is limited. This results in the question of how to present virtual experts while preserving the valuable elements of in-person demonstration.

\textbf{Hybrid} solutions also exist using synchronous and asynchronous technology where experts can record and stream XR instructions~\cite{nebelingXRStudioVirtualProduction2021} or work on simulations~\cite{dossantosCollaborativeVRVisualization2011} at different but also at the same time. In the following, we will refer to asynchronous applications due to the specific use case, but we will not exclude that the results can also be used for synchronous developments. 

\subsection{Visualization}
As outlined above, we saw that current synchronous and asynchronous research mostly pays attention to the technical aspects of explanations and less to the social and emotional components that impact the experience. We learned that factors such as demonstration and imitation, but also embodiment and social connectedness, play an important role in learning success and can offer additional value if applied correctly. By changing the perspective and providing gaze cues for the learner, these aspects can be influenced. In the following, we will discuss these factors in more detail.

\textbf{Perspective} change is a commonly used tool in video production to get into a different viewpoint. 2D tutorials and vlogs change the camera angle to see what another person sees, the so-called 1PP. This perspective outperforms the 3PP in practical instructions~\cite{fiorellaItAllMatter2017}. However, the effectiveness of this approach is also dependent on the learners' handedness~\cite{kellyDifferentialMechanismsAction2013}. As perspective is rather a continuum in the three-dimensional space~\cite{hoppeThereNoFirst2022}, research has explored different viewpoints for asynchronous~\cite{thanyaditXRLIVEEnhancingAsynchronous2022} Virtual Reality (VR) and synchronous XR usage~\cite{chenechalVishnuVirtualImmersive2016}. When it comes to embodiment, the feeling towards the body can change with the switch from 3PP to 1PP, but even in 3PP, there is still a feeling of embodiment possible~\cite{liouDistinctionFirstpersonPerspective2023}. This feeling of embodiment while being in the 1PP of an expert can help to achieve better performance~\cite{banakouVirtuallyBeingEinstein2018} and connection towards the other~\cite{yongChangeSceneryTransformative2024}. When we look at already implemented asynchronous VR instructions, reduced ghost/shadow or solid instructor visualizations are used to preserve the orientation of the expert. Thanyadit et al.~\cite{thanyaditXRLIVEEnhancingAsynchronous2022} compared these to videos with different viewpoints, whereby some participants even tried to take on the instructor's perspective in the shadow variant. This leads to the participant getting confused because of the overlapping of virtual objects. Synchronous Augmented Reality (AR) research has already explored displaying the students' actions in 1PP or 3PP to the expert~\cite{sunEmployingDifferentViewpoints2018} while asynchronous AR work varied different visualizations of the 3PP instructor~\cite{caoExploratoryStudyAugmented2020}, implemented a ghost 1PP visualization~\cite{hanARArmAugmentedVisualization2016}, or only the tools' movement in the recordings~\cite{chidambaramProcessARAugmentedRealitybased2021}. This means we have not found research using the experts' asynchronous 1PP and comparing it to 3PP in AR. However, the previously described research shows that perspective has an important influence on how well we can understand manual explanations. While 1PP gives us a better view and feeling towards a task, the 3PP visualizes more important cues, showing the instructor's attention. It is therefore unclear how these findings generalize to our use case.

\textbf{Gaze} is an important social cue influencing memory, attention~\cite{contyLookMeLl2012}, sense of being together, and communication~\cite{piumsomboonEffectsSharingAwareness2019a}. Additionally, it can also indicate future actions~\cite{zhengEgocentricEarlyAction2023} of a person and remains available when both hands are in use. Gaze can be divided into eye and head gaze, with eye gaze showing the focus of the eyes and head gaze visualizing the direction of the head. While head gaze is slower and not as precise as eye gaze, it is easier to measure as it needs less technology, less workload, calibration, and signals higher interest~\cite{sidenmarkEyeHeadSynergetic2019}. Moreover, if the position of the head is not fixed, the eye gaze also depends on the direction of the head to track the focus in the environment~\cite{sidenmarkEyeHeadTorso2020}.
For joint gaze, the synchronously shared gaze of an expert can improve the learner's effectiveness, especially for novices, conveying content-related and procedural information for visuospatial tasks~\cite{sungLearnersLearnMore2021}. Additionally, research shows that the gaze of an expert is more focused compared to a novice one while solving a task~\cite{yangAttentionDynamicsProgramming2024a}. XR developments use gaze cues for analysis~\cite{siewImprovingMaintenanceEfficiency2019} and the adaptation of virtual tutors~\cite{khokharModifyingPedagogicalAgent2022}. Gaze can be depicted for small focus areas with dots~\cite{sungLearnersLearnMore2021} and beams~\cite{piumsomboonEffectsSharingAwareness2019a} as they mainly display one specific point. Larger gaze areas like head gaze can be visualized with cones~\cite {piumsomboonEffectsSharingAwareness2019a} and pyramid frustums~\cite{piumsomboonEffectsSharingAwareness2019a}. For both and shared gaze in particular ovals~\cite{bovoConeVisionBehavioural2022}, circles~\cite{jingEyeSeeWhat2021}, trajectory, highlights, and spotlights~\cite{zhangLookTogetherUsing2017} are used. Circular cursors are the most common option for all kinds of areas, and for example also used as simplified cone frustums on 2D surfaces~\cite{bovoConeVisionBehavioural2022}. When it comes to social factors, synchronous XR head-based~\cite{piumsomboonEffectsSharingAwareness2019a, bovoConeVisionBehavioural2022} as well as eye-based cues can enhance collaboration and social connectedness~\cite{jingEyeSeeWhat2021}. Still, we could not find research on the influence of gaze visualizations on the social aspects in asynchronous XR instruction scenarios, nor between 1PP and 3PP for XR in general.

In summary, current research highlights the role of social and embodiment factors, emphasizing the importance of demonstrations, social connectedness, and cues in knowledge transfer. XR technologies offer synchronous and asynchronous methods for enhancing practical instructions. While asynchronous approaches provide flexibility and accessibility, the social and emotional aspects are often overlooked. In particular, visualization techniques, such as perspective shifts and gaze cues, seem promising for improving understanding and engagement, still their application in asynchronous XR, especially AR, remains limited. Bridging these gaps can enhance the effectiveness of future knowledge transfer.

\section{Methodology}
 Previous work demonstrated that asynchronous XR instructions can support explanations that need to be delivered effectively and accurately, regardless of time. However, our analysis of related work revealed that the influence of perspective and important social indicators, such as gaze cues in AR, has not yet been thoroughly investigated. Following the work in related areas, these aspects can improve productivity and the learning experience in synchronous scenarios and other media. In addition, switching to another person's perspective also changes factors such as the embodiment and social connectedness to that person, which in turn impacts the learning outcome and experience. This emphasizes the importance of the question of how these aspects influence each other during asynchronous XR, especially AR instructions. With the approval of an ethics committee, we investigated these factors and conducted a user study to answer the following research questions:

\begin{enumerate}
    \item[\textbf{RQ1}] How does the \textit{perspective} affect the efficiency, embodiment, and social connectedness during asynchronous instructions of two-handed manual tasks?
    \item[\textbf{RQ2}] How does the \textit{gaze} affect the efficiency, embodiment, and social connectedness during asynchronous instructions of two-handed manual tasks?
    \item[\textbf{RQ3}] Are there \textit{interaction effects} between the perspective and gaze on efficiency, embodiment, and social connectedness during asynchronous instructions of two-handed manual tasks?
\end{enumerate}

\subsection{Study Design}
\begin{figure}
    \centering
    \includegraphics[width=\linewidth]{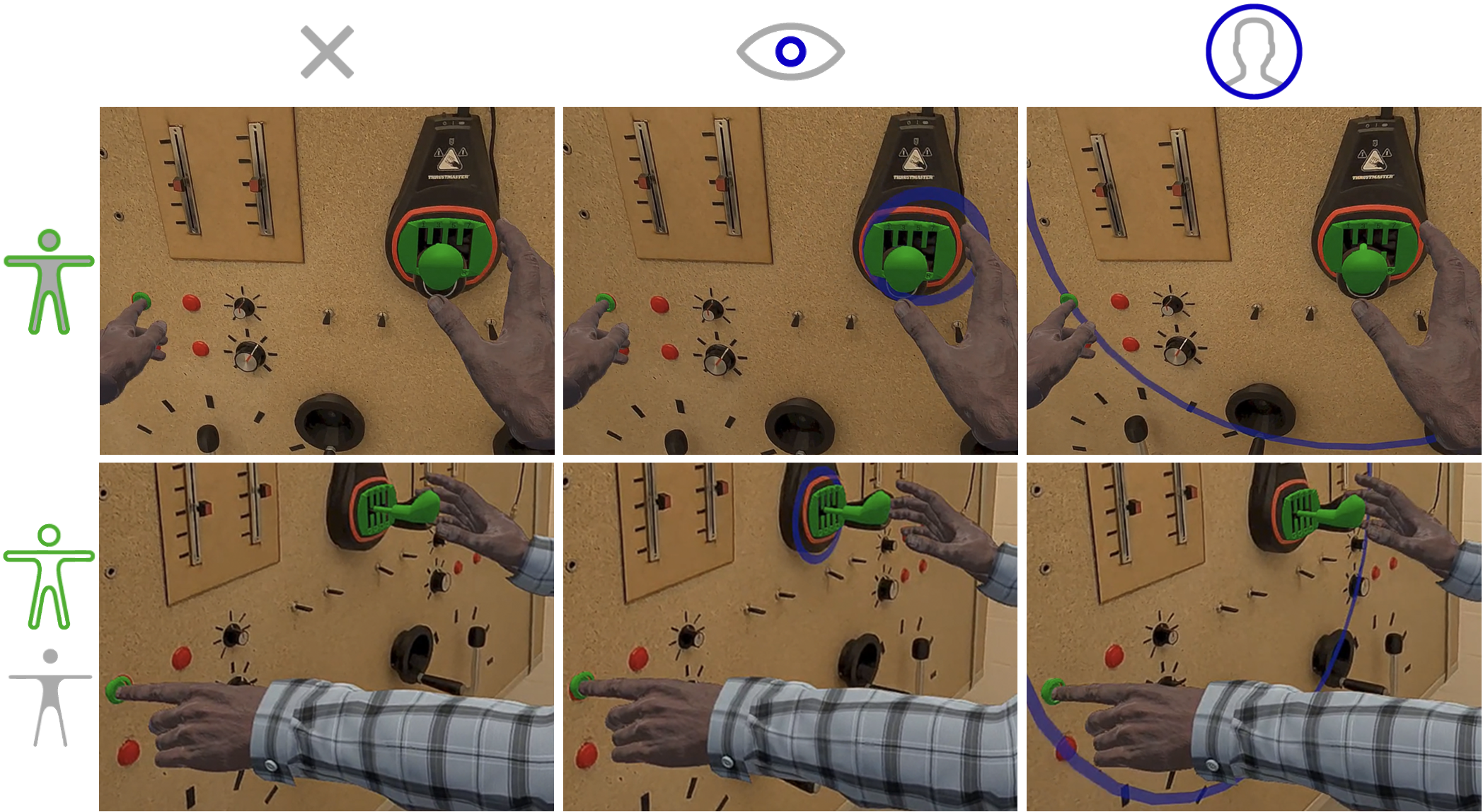}
    \caption{\textit{1PP} and \textit{3PP} \textit{perspective} as well as \textit{none}, \textit{eye} and \textit{head} \textit{gaze} conditions during the asynchronous two-hand machine task instructions in AR.}
    \label{fig:conditions}
\end{figure}

To answer these research questions, we designed a user study with the \textit{perspective} and \textit{gaze} cues of the asynchronous instructor as the independent variables. For the first independent variable \textit{perspective}, we varied the two levels \textit{1PP} and \textit{3PP} of the instructor to analyze the influence on efficiency, embodiment, and social connectedness (Figure~\ref{fig:teaser}). For the second independent variable \textit{gaze}, we investigated the effects of the three levels \textit{none}, \textit{head}, and \textit{eye} gaze on the above-mentioned aspects. We varied both independent variables in a within-subjects design with a total of $2\times3 = 6$ conditions (Figure~\ref{fig:conditions}) and counterbalanced the order using a balanced Latin square.

\subsection{Task}
To evaluate the efficiency of the imitations precisely, we designed the task based on other asynchronous XR publications that used a machine interface~\cite{caoExploratoryStudyAugmented2020, liuInstruMentARAutoGenerationAugmented2023}. 
In addition, we decided on body-coordinated tasks to have a suitable and dynamic use case for the different gaze visualizations. Each interaction involved both the left and right hands during an explanation step. With a focus on only the demonstration and imitation as the crucial initial phases of learning, as well as the visual aspects of the explanations, the virtual expert showed a task to the user on an interface. The user then had to repeat the interaction after the animation ended. Depending on the condition, the instructor was visualized in \textit{1PP} with only the hands or in \textit{3PP} with the whole body. During the animations, the users had to change their position either to where the expert is located (\textit{1PP}) or beside the expert (\textit{3PP}). In \textit{3PP}, the user had to go to the position of the expert after the animation ended to start the interface interaction. This is the same process as if a real person were present on-site, giving machine interface instructions and afterward handing over to the learner. Depending on the condition, the system added \textit{none}, \textit{eye}, or \textit{head} gaze cues to potentially give additional guidance during the explanation. 

\begin{figure}
    \centering
    \includegraphics[width=0.732\linewidth]{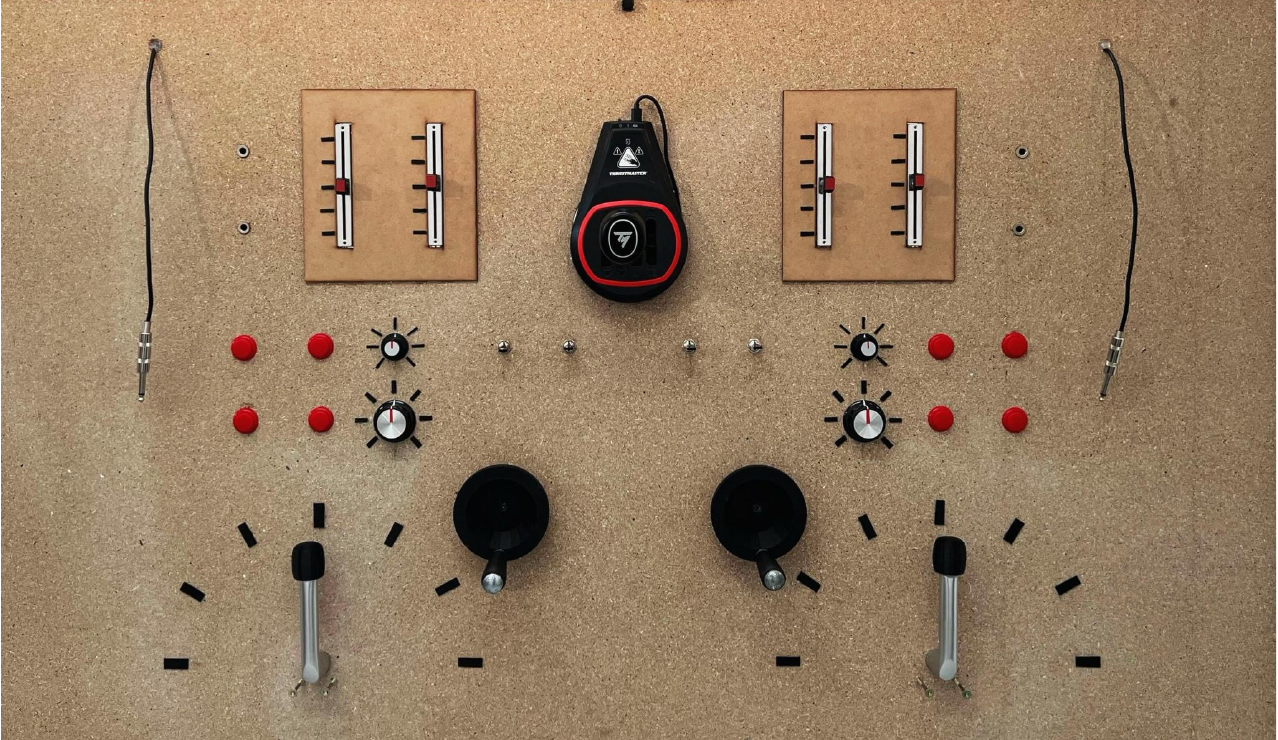}
    \caption{Arrangement of the individual elements on the machine interface consisting of the same amount of pins, buttons, sliders, knobs, control switches, wheels, and levers on both sides, as well as a shift and stop button in the middle.}
    \label{fig:interface}
\end{figure}

We based parts of the study design and the machine interface on the work of Cao et al.~\cite{caoExploratoryStudyAugmented2020} and added additional elements that are common in music production, electrical engineering, craftsmanship, cars, and airplanes. Therefore, we implemented multiple elements: 8 buttons, 4 switches, 2 levers, 4 knobs, 4 sliders, 2 wheels, and 4 pin sockets, including 2 pins, which we mounted mirrored on the left and right of a vertical wood panel. To make them more reachable, we added 1 shift and a stop button in the top middle of the panel that the users had to press after finishing a run. The construction is shown in Figure~\ref{fig:interface}.
We designed a sequence of 12 different element combinations for both hands (1:switch-knob, 2:button-shift, 3:slider-slider, 4:button-wheel, 5:pin-switch, 6:shift-wheel, 7:knob-slider, 8:lever-lever, 9:pin-button, 10:switch-shift, 11:pin-knob, 12:wheel-lever) that included every element type 3 times. To avoid learning effects, we repeated these 12 combinations 6 times using different values and elements from the same element type to keep the 6 conditions comparable. This results in a total of $12\times6 = 72$ combinations. While only varying the order of our 6 conditions, the 72 sequences, values, and elements stayed the same for every participant. For example, a participant experienced the first animation in a combination of the first switch on the left, the knob on the bottom right, \textit{1PP}, and \textit{none} while the next participant also had to interact with the same switch on the left and knob on the bottom right first, but sees the \textit{3PP} and \textit{head} condition instead.

\subsection{Dependent Variables}
To answer the research questions, we collected data using the loggers of the machine interface, videos, quantitative and qualitative questionnaires, as well as conducted interviews, resulting in the following dependent variables during each condition:

\textbf{Accuracy:} The value of successfully completing the task, measured by logging all the machine interface's electronic parts and reviewed by two authors through the recorded videos. We used different criteria for estimation, like the correct value adjustment, correct element order, correct hand usage, and the combined overall correct execution. For elements with continuous values, we measured and calculated the boundary between two setting options, resulting in reference areas for the correct value adjustment. Additionally, we estimated the correct hand usage through the video material.

\textbf{Task Completion Time:} The time the participants needed to complete the task, calculated with the logger data in two ways using the beginning of the animation (B-S) or the first interaction with the machine interface (E-S) as the start and pressing the stop button as the end input. We further estimated if participants waited to start the task before the animation ended (Patience).

\textbf{Mental Load:} An efficiency influencing factor evaluated using the RAW (NASA-)TLX~\cite{hartDevelopmentNASATLXTask1988} (21-point Likert scale, 0:~Very Low, 20:~Very High).

\textbf{Task Difficulty:} The manual tasks' difficulty that influences the efficiency using the Single Ease Question (SEQ)(7-point Likert scale, 1:~Very Difficult, 7:~Very Easy).

\textbf{Social Connectedness:} The feeling towards the instructor using the Inclusion of Other and Self (IOS)~\cite{aronInclusionOtherSelf1992a} (7-point Likert scale visualized through circles with 1 being not at all close and 7 being extremely close) to evaluate the perceived closeness to the instructor as well as the Game Experience Questionnaire Social Presence Modules Empathy (GEQ-SPM-E) and Behavioural Involvement (GEQ-SPM-BI) components~\cite{ijsselsteijnGameExperienceQuestionnaire2013b} (5-point Likert scale, 1: not at all, 5: extremely).

\textbf{Embodiment:} The feeling of embodiment towards the virtual expert using the short Avatar Embodiment questionnaire~\cite{peckAvatarEmbodimentStandardized2021a} (7-point Likert scale, 1:~Strongly disagree, 7:~Strongly agree).

\textbf{Instructions:} Our own questions about how the participants perceived the instructions (7-point Likert scale, 1:~Strongly disagree, 7:~Strongly agree).

\textbf{Performance:} Our own questions regarding the participants' and instructors' performance (7-point Likert scale, 1:~Strongly disagree, 7:~Strongly agree).

Additionally, we wrote notes and also gave the opportunity to leave written comments at the end of the questionnaire after each condition.
The semi-structured interview after the main study revolved around the elements, \textit{perspective}, \textit{gaze} cues, instructions, instructor, task difficulty, and general suggestions. We used a mixed-methods approach to get insight into the experiences of the participants, especially regarding embodiment and social connectedness.

\begin{figure}
    \centering
    \includegraphics[width=\linewidth]{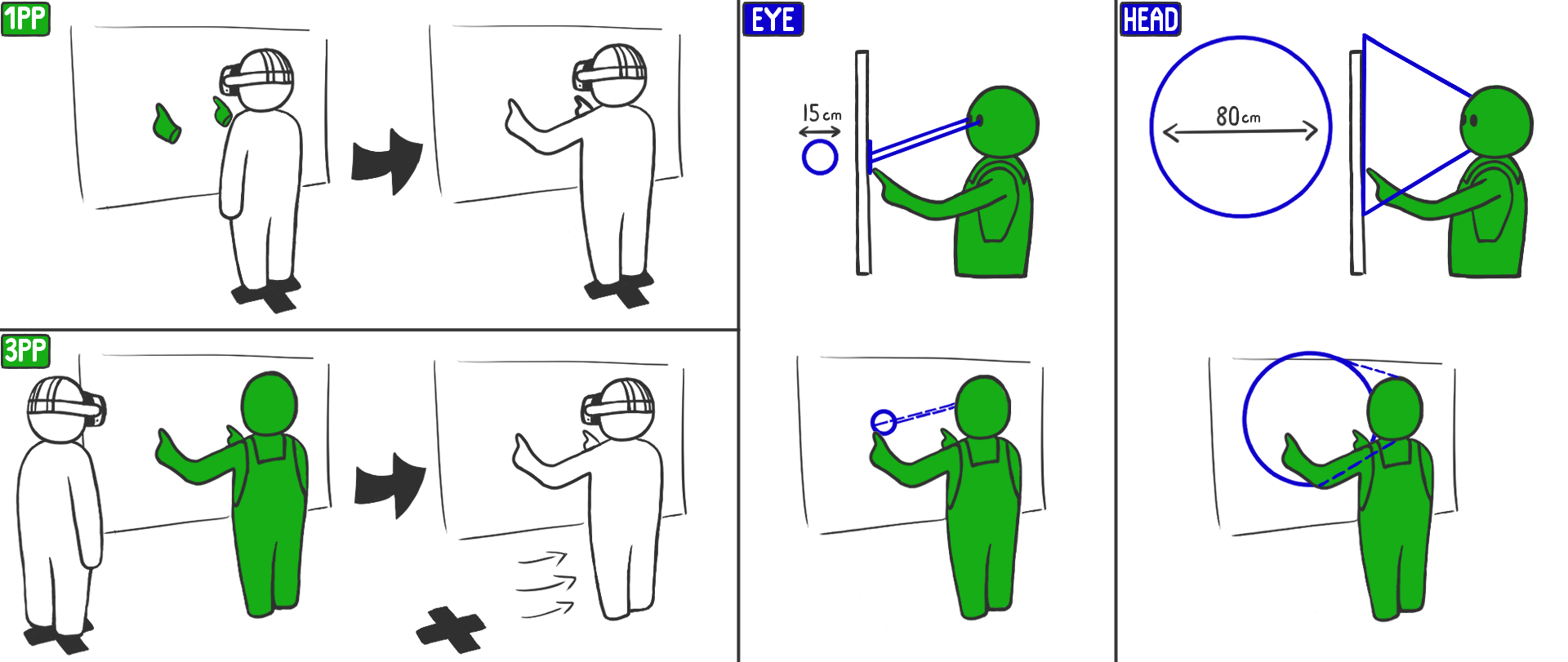}
    \caption{Step sequences of the user during the different perspectives (left) and visualization of the blue gaze cues of the AR expert (right). The cross on the floor marks the standing position of the user while watching the instructions.}
    \label{fig:step}
\end{figure}

\subsection{Apparatus}
Based on these considerations, we designed a machine interface and an AR application. We build the interface using a 3D printer, jigsaw, laser cutter, drill, milling machine, soldering iron, and hot glue. To measure the interactions of all elements except the shift, we used sensors like buttons, switches, sliders, rotary encoders, and potentiometers. We coupled these to an Arduino Mega ADK connected to a PC. The industrially produced shift was linked directly to a PC without an intermediate microcontroller. To implement the loggers, we used the Arduino IDE and Python.

Further, we created the AR application with Unity. For the expert, we decided on a full-body realistic avatar, which was the most fitting for the embodiment and social connectedness evaluation while blending in with the environment. Using Character Creator 4 for the avatar, we emphasized friendly features, detailed hands, a complete face, and body rigging as well as elaborated blend shapes. With the study supervisor identifying as female, we decided on a male-looking character to keep the experiment more diverse. In addition, the full-body avatar was also rated high, especially for body-coordinated tasks in previous asynchronous AR works~\cite{caoExploratoryStudyAugmented2020}. Using an avatar made it possible to capture the \textit{head} and \textit{eye gaze} with the Head-Mounted Display (HMD) compared to volumetric videos.  The instructor avatar wears a craftsman's overall as the clothing can influence the personality perception, including competence or achievement~\cite{legdeEvaluatingEffectClothing2019}.

We recorded the avatar interaction and \textit{gaze} with the Meta Quest Pro using the positional and rotational data for the \textit{head} \textit{gaze}, the eye tracking feature for the \textit{eye} gaze, as well as the inside-out body and hand tracking. Due to the lack of asynchronous XR gaze cue research, we decided on a blue circular cursor~\cite{jingImpactSharingGaze2022, piumsomboonEffectsSharingAwareness2019a} for both cues with each having a different size, to keep the design of the \textit{gaze} cues consistent and comparable to synchronous related work. For the \textit{head} \textit{gaze} we estimated and prior tested the size of the circle using the maximum 106° horizontal field of view of the recording HMD as reference for our calculations~\cite{piumsomboonEffectsSharingAwareness2019a, sidenmarkEyeHeadSynergetic2019} to make them visible, less irritating, not cover elements, the \textit{eye gaze} cues more comparable, and include all the perceived elements of the instructor as the \textit{eye}-\textit{head} coordination can highly vary between people~\cite{sidenmarkEyeHeadTorso2020}. With an average distance of 30~cm between the expert and the machine interface, this leads to a diameter of approximately 80~cm. For the \textit{eye} \textit{gaze} cue, we used a diameter of 15~cm to not obscure the focused elements during the animation and symbolize the joint viewing state of the participant and instructor~\cite{jingImpactSharingGaze2022}. As the virtual avatar can not change the real environment, we created digital twins in green for each element and fitted them to the avatar recordings. These are often used in remote applications~\cite{odaVirtualReplicasRemote2015} to make the explanations more realistic and clear. Additionally, the length of each animation was normalized between the 6 conditions to an average length of 6.92 seconds per animation so that they are more comparable~\cite{caoExploratoryStudyAugmented2020}. We added real markers on the floor to label the two positions the participant has to take, depending on the condition, in combination with virtual ones that showed the current location to stand during the animation. Visualizations of the position change and the \textit{gaze} cues are shown in Figure~\ref{fig:step}.

For the study, we used the Meta Quest 3 due to its clearer pass-through and kept it always connected to a PC for study control and panel element adjustments. This was necessary because, at the time of the study, the HMD only allowed semantic classification labels like room architecture and furniture. An anchor near the panel was adjusted before each run.

In order to measure the dependent variables' accuracy and task completion time, we recorded the timestamp in ms, the value, and the element name if the value of a machine part changed. We also logged when an anchor was set in AR, the supervisor pressed the next animation start button on the PC, the animation ended, or the animation was replayed. In addition, we recorded videos of the interface during the study, which we used to verify the variables and estimate if the participants used both hands during the interaction. 

\begin{figure}[tbp]
  \centering
  \begin{subfigure}[b]{0.68\columnwidth}
  	\centering
  	\includegraphics[width=\textwidth]{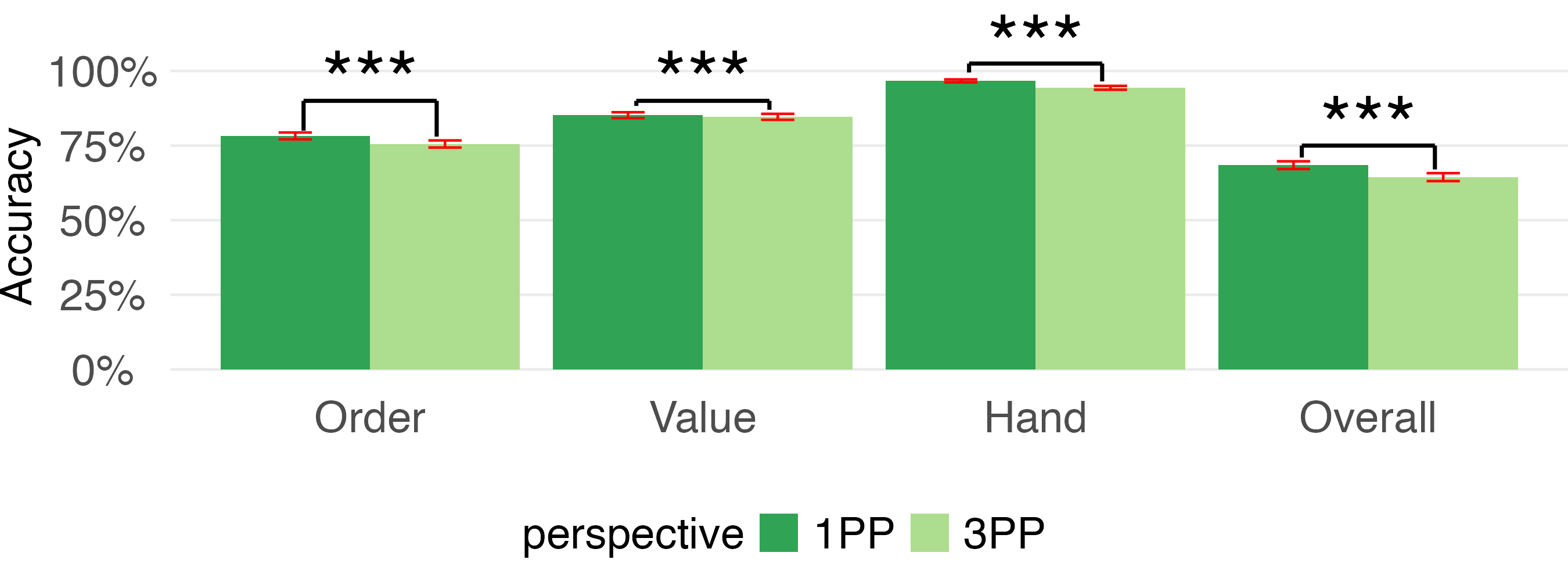}
  	\caption{Accuracy}
  	\label{fig:results:performance:accuracy}
  \end{subfigure}%
  \hfill%
  \begin{subfigure}[b]{0.32\columnwidth}
  	\centering
  	\includegraphics[width=\textwidth]{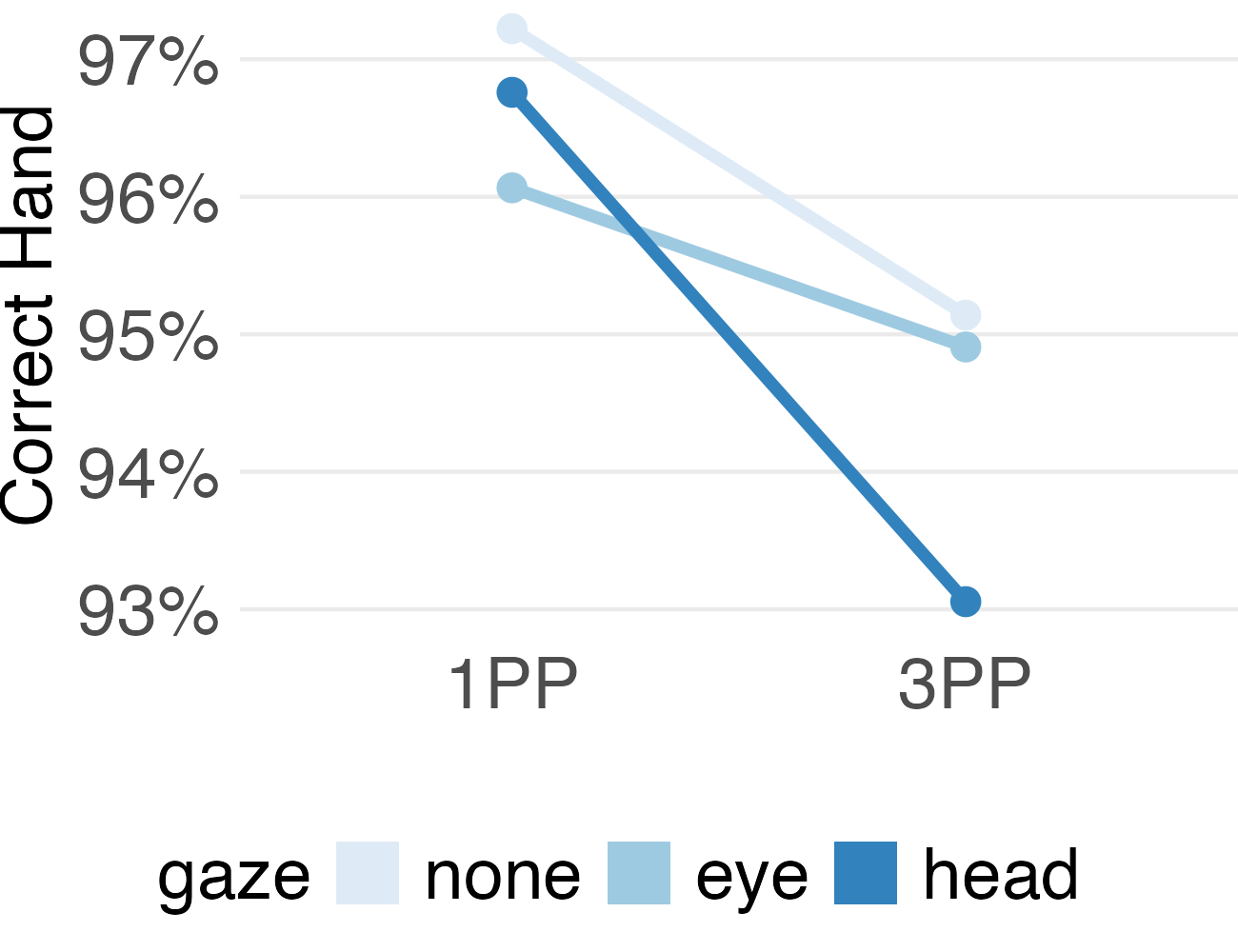}
  	\caption{Correct Hand}
  	\label{fig:results:performance:hand}
  \end{subfigure}%
  \\%
  \caption{Bar chart of the Accuracy by \textit{perspective}~(\subref{fig:results:performance:accuracy}) as well as bar and interaction charts of the correct hand by \textit{perspective}$\times$\textit{gaze}~(\subref{fig:results:performance:hand}). The error bars indicate the standard error.}
  \label{fig:accuracy}
\end{figure}

\subsection{Procedure}
After welcoming the participants, we introduced them to the concept and asked them to fill out a consent form and a pre-questionnaire including demographic data and prior knowledge. After they put on the glasses, we started the application and jointly adapted the virtual machine elements to cover the real ones.

To begin the first condition, we guided the participants to one of the virtual markers on the floor and started the first animation as soon as they were ready. After the animation ended, the participants should, depending on the condition, go to or stay in the position of the virtual expert. They could then start adjusting the elements following the example of the virtual instructor. By pressing the stop button and saying that they completed the task, the participants signaled the end of the step. If an element was set incorrectly that was used later on in the condition, we asked the participants to correct it and push the stop button again. We then started the next animation by using the keyboard on the computer. This process happened a total of 12 times before the participants took the HMD off and answered a questionnaire about the condition on a PC. This break reduced the risk of them getting cybersickness. After the participants finished the questionnaire, the whole process started again in the same sequence with different conditions, elements, and values. 

After the participants completed all 6 conditions, we conducted a short interview about their experience, which we audio recorded. Each experiment took about 80 minutes per participant.

\subsection{Participants}
We recruited 36 participants (19 male, 16 female, 1 non-binary), aged between 19 and 68 ($Mean = 31.31$, $SD = 13.28$). 33 of the participants were right- and 3 left-handed, while the height varied between 1.57 m and 1.93 m. Among the participants were 26 students, two designers, two managers, one editor, one architect, one care worker, one former recruiter, one secretary, and one social worker. All participants voluntarily took part in the study and got reimbursed 10€ per hour or study points. 

\subsection{Analysis}
We analyzed the study data using \textit{Linear Mixed-Effects Models} (LMM) and tested them for normal distribution using Quantile-Quantile (Q-Q) plots by Wilk~and~Gnanadesikan~\cite{wilkProbabilityPlottingMethods1968}. If the data showed no normality, we performed an \textit{Aligned Rank Transformation} (ART) as proposed by Wobbrock~et~al.~\cite{wobbrockAlignedRankTransform2011a}. Further, we analyzed count values with \textit{generalized linear mixed models TMB} (glmmTMB) as described by Brooks~et~al.~\cite{brooksGlmmTMBBalancesSpeed2017} in combination with \textit{Poisson} regression model. For an estimate of the mean response for normally distributed data, we report the \textit{Estimated Marginal Mean} (EMM) by Searle~et~al.~\cite{searlePopulationMarginalMeans1980} using the \textit{Bonferroni} correction. If the data showed no normality, we performed \textit{post-hoc} tests by Elkin~et~al.~\cite{elkinAlignedRankTransform2021a}. For all plots, we highlighted significant differences ($***:$ p$<$.001, $**:$ p$<$.01, and $*:$ p$<$.05).

\section{Results}
In the following, we report the results of our controlled user study to answer our research questions. In the pre-questionnaire, the participants rated their prior experience with AR ($Median = 3$), HMD ($Median = 5$), and machine interfaces ($Median = 5$) on a 7-point Likert scale, with 1 having no experience and 7 being an expert.

\subsection{Accuracy}
We determined the accuracy by analyzing the participants' correct value adjustments, element order, and hand usage. The analysis using ART shows that the \textit{perspective} influences the correct order (F$_{1,2551}$ = 31.43, p$<$.001), value of the elements (F$_{1,2551}$ = 19.25, p$<$.001) as well as the hand (F$_{1,2551}$ = 615.45, p$<$.001) resulting in an effect on the overall correctness (F$_{1,2551}$ = 147.80, p$<$.001) with \textit{1PP} being the better option as can be seen in Figure~\ref{fig:results:performance:accuracy}. Additionally, we found an effect between \textit{perspective} and \textit{gaze} for the correct hand (F$_{1,2551}$ = 3.31, p$<$.05) with the post-hoc test showing a significant effect between (\textit{1PP},~\textit{none}) and (\textit{3PP},~\textit{head}) (p$<$.05) with the first one being more accurate (see Figure~\ref{fig:results:performance:hand}). 

\subsection{Task Completion Time}
\begin{figure}[tbp]
  \centering
  \begin{subfigure}[b]{0.25\columnwidth}
  	\centering
  	\includegraphics[width=\textwidth]{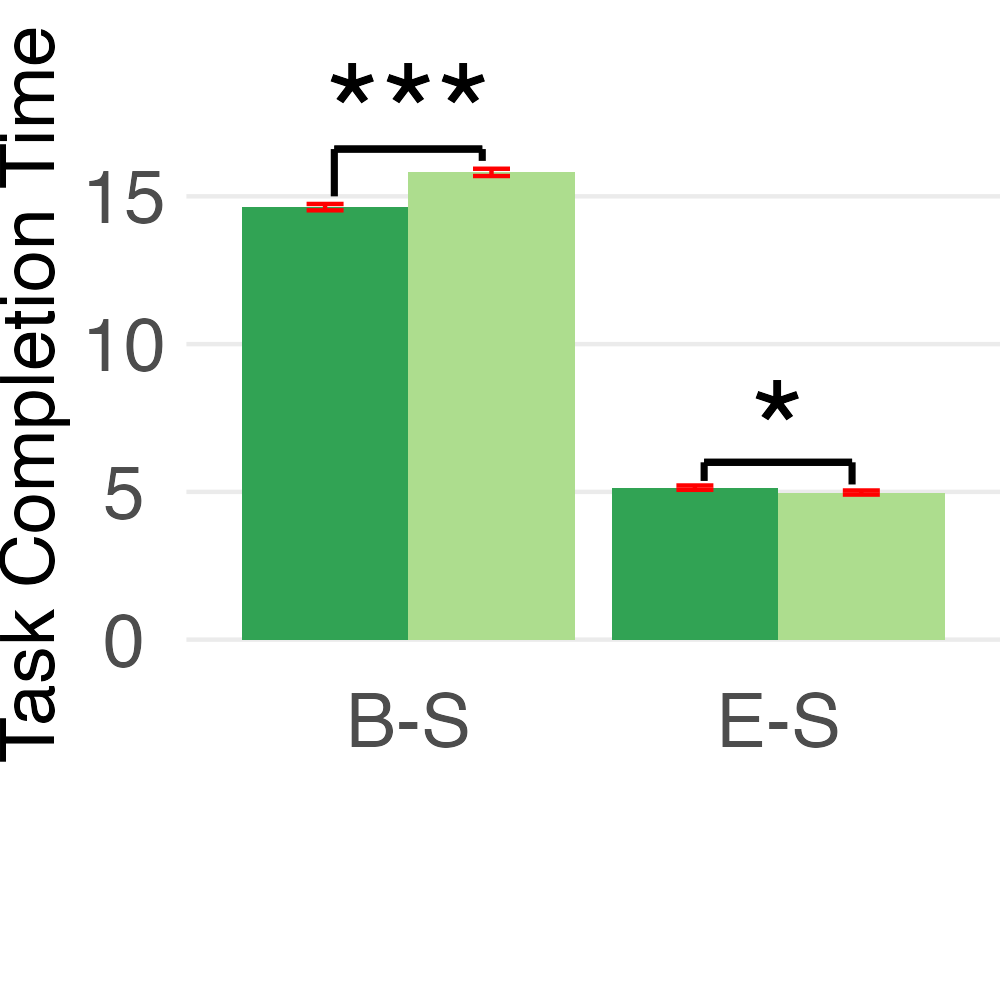}
  	\caption{Time in s}
  	\label{fig:results:performance:time}
  \end{subfigure}%
  \hfill%
  \begin{subfigure}[b]{0.25\columnwidth}
  	\centering
  	\includegraphics[width=\textwidth]{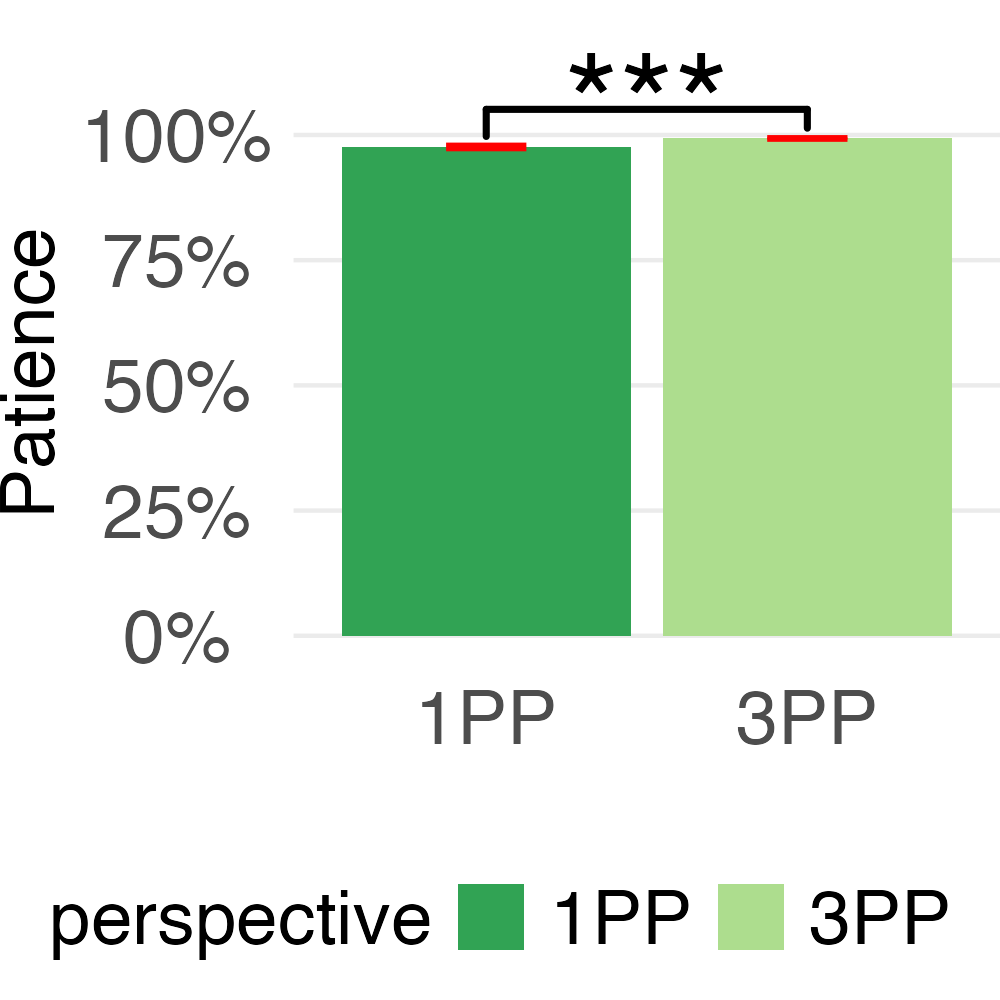}
  	\caption{Patience}
  	\label{fig:results:performance:patienceP}
  \end{subfigure}%
  \begin{subfigure}[b]{0.25\columnwidth}
  	\centering
  	\includegraphics[width=\textwidth]{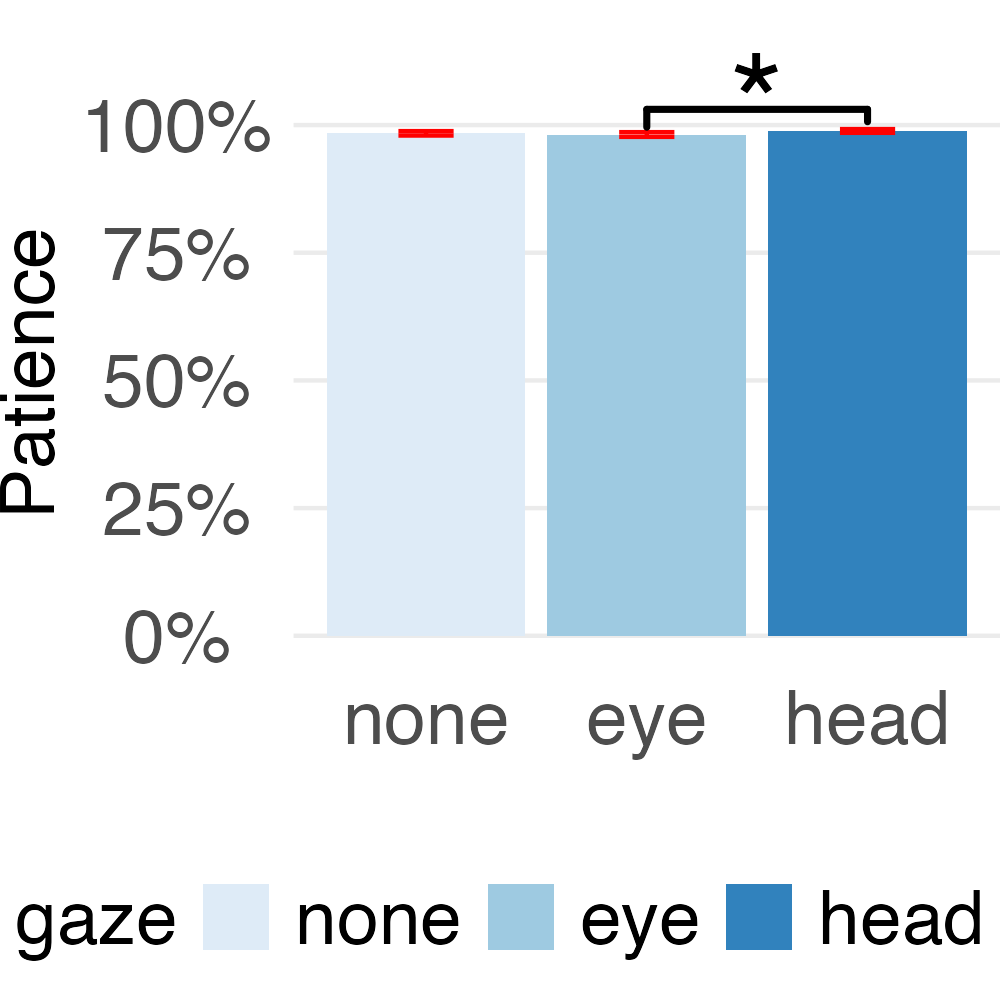}
  	\caption{Patience}
  	\label{fig:results:performance:patienceG}
  \end{subfigure}%
  \begin{subfigure}[b]{0.25\columnwidth}
  	\centering
  	\includegraphics[width=\textwidth]{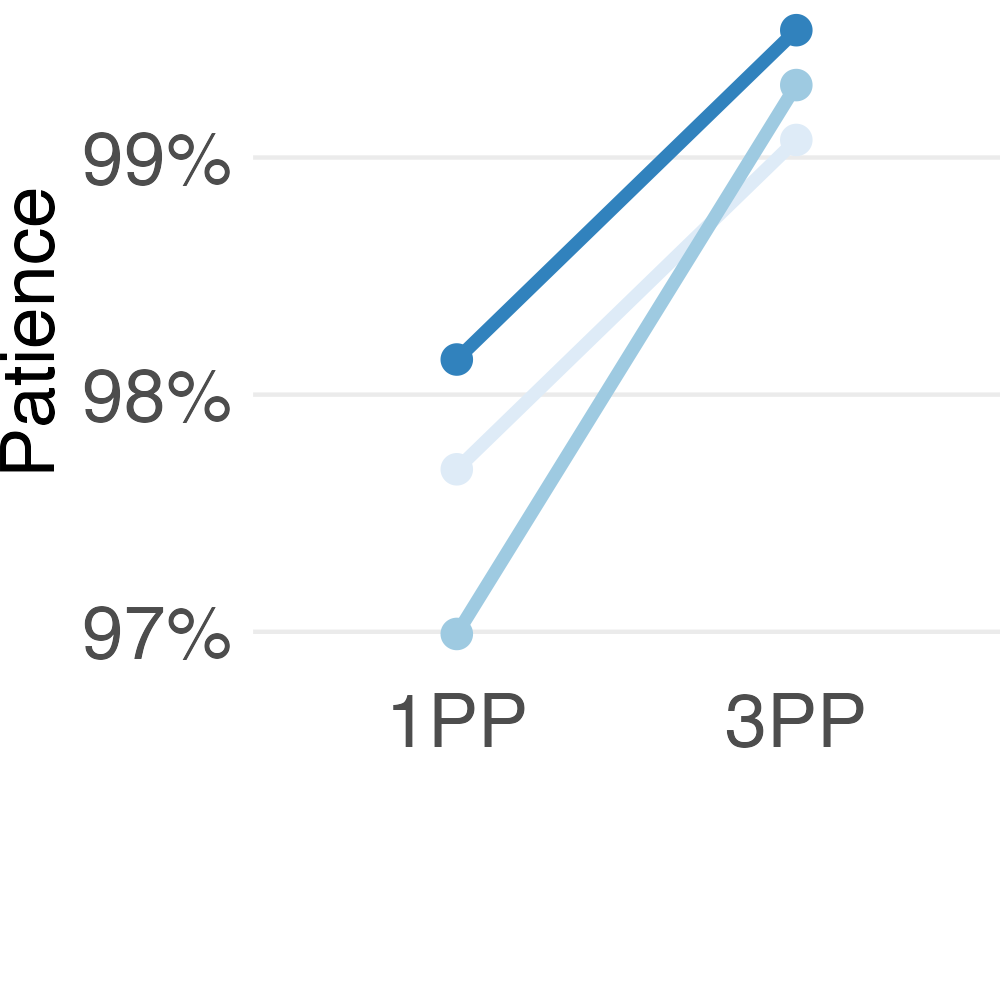}
  	\caption{Patience}
  	\label{fig:results:performance:patiencePG}
  \end{subfigure}%
  \\%
  \caption{Bar chart of the Task Completion Time by \textit{perspective}~(\subref{fig:results:performance:time}) as well as bar and interaction charts of the Patience by \textit{perspective}~(\subref{fig:results:performance:patienceP}), \textit{gaze}~(\subref{fig:results:performance:patienceG}) and \textit{perspective}$\times$\textit{gaze}~(\subref{fig:results:performance:patiencePG}). The error bars indicate the standard error.}
  \label{fig:enter-label}
\end{figure}

We assessed the Task Completion Time in two ways: the beginning of the animation until the press of the stop button (B-S) and the first interaction with a machine element until the press of the stop button (E-S). Additionally, we tracked the Patience of the participants and wanted to replay the instructions, as these could influence the interaction length. The ART analysis shows that the \textit{perspective} significantly affects both the B-S (F$_{1,2551}$ = 106.69, p$<$.001) and E-S (F$_{1,2551}$ = 5.25, p$<$.05) with the time being higher in the \textit{3PP} for the B-S and higher in the \textit{1PP} for the E-S as can be seen in Figure~\ref{fig:results:performance:time}. Further, the analysis using ART shows that the \textit{perspective} influences the Patience (F$_{1,2551}$ = 522.06, p$<$.001) with \textit{3PP} being the more patient one (see Figure~\ref{fig:results:performance:patienceP}). Also, the \textit{gaze} affects the earlier interaction (F$_{1,2551}$ = 3.11, p$<$.05) with the post-hoc test confirming that the participants were more patient in the \textit{head} compared to \textit{eye} condition (p$<$.05) (see Figure~\ref{fig:results:performance:patienceG}). Moreover, the analysis shows interaction effects of \textit{perspective} and \textit{gaze} on the Patience (F$_{1,2551}$ = 3.24, p$<$.05). The post-hoc test reveals that (\textit{1PP},\textit{eye}) was the more impatient condition in comparison to (\textit{3PP},\textit{head}) (p$<$.01) or (3PP,\textit{eye}) (p$<$.05) as can be seen in Figure~\ref{fig:results:performance:patiencePG}. The results of the replayed animations showed no significant differences ($Mean = 0.04$) using glmmTMB. 

\subsection{Mental Load}
\begin{figure}[tbp]
  \centering
  \begin{subfigure}[b]{0.25\columnwidth}
  	\centering
  	\includegraphics[width=\textwidth]{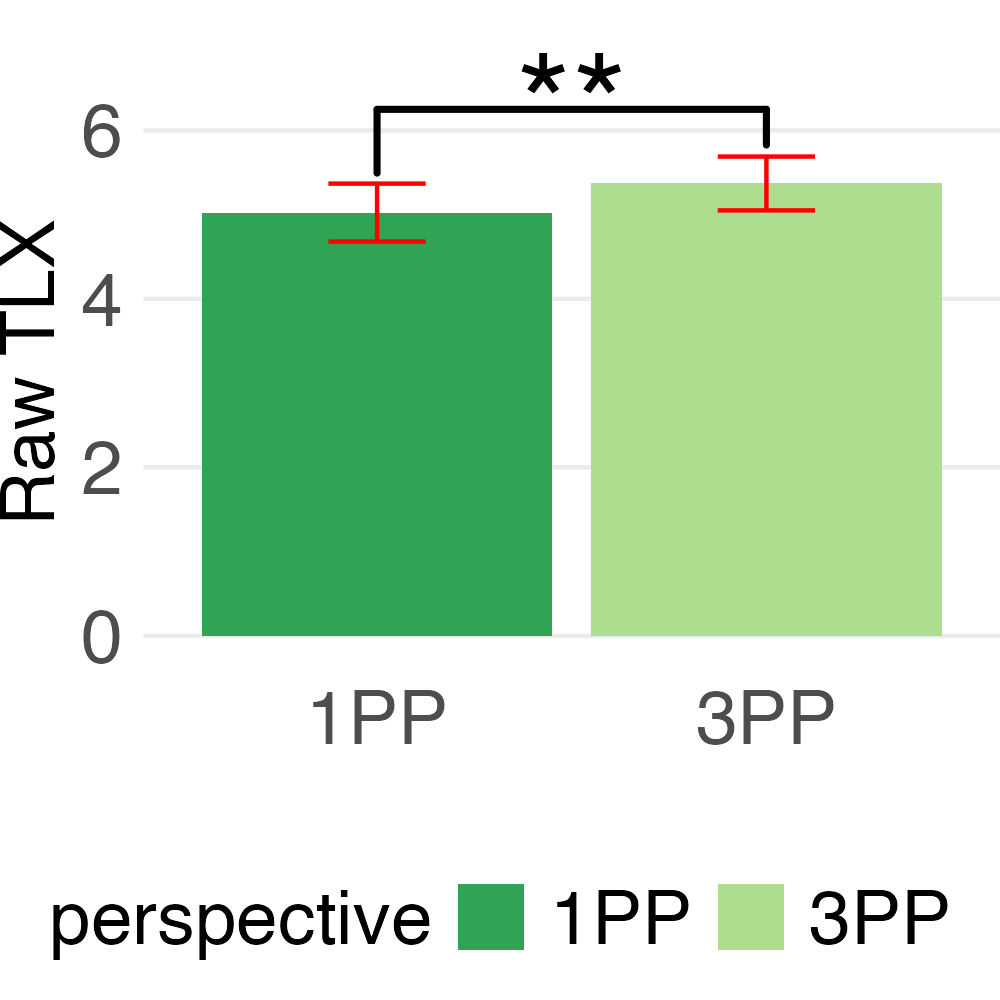}
  	\caption{Raw TLX}
  	\label{fig:results:performance:tlxP}
  \end{subfigure}%
  \begin{subfigure}[b]{0.25\columnwidth}
  	\centering
  	\includegraphics[width=\textwidth]{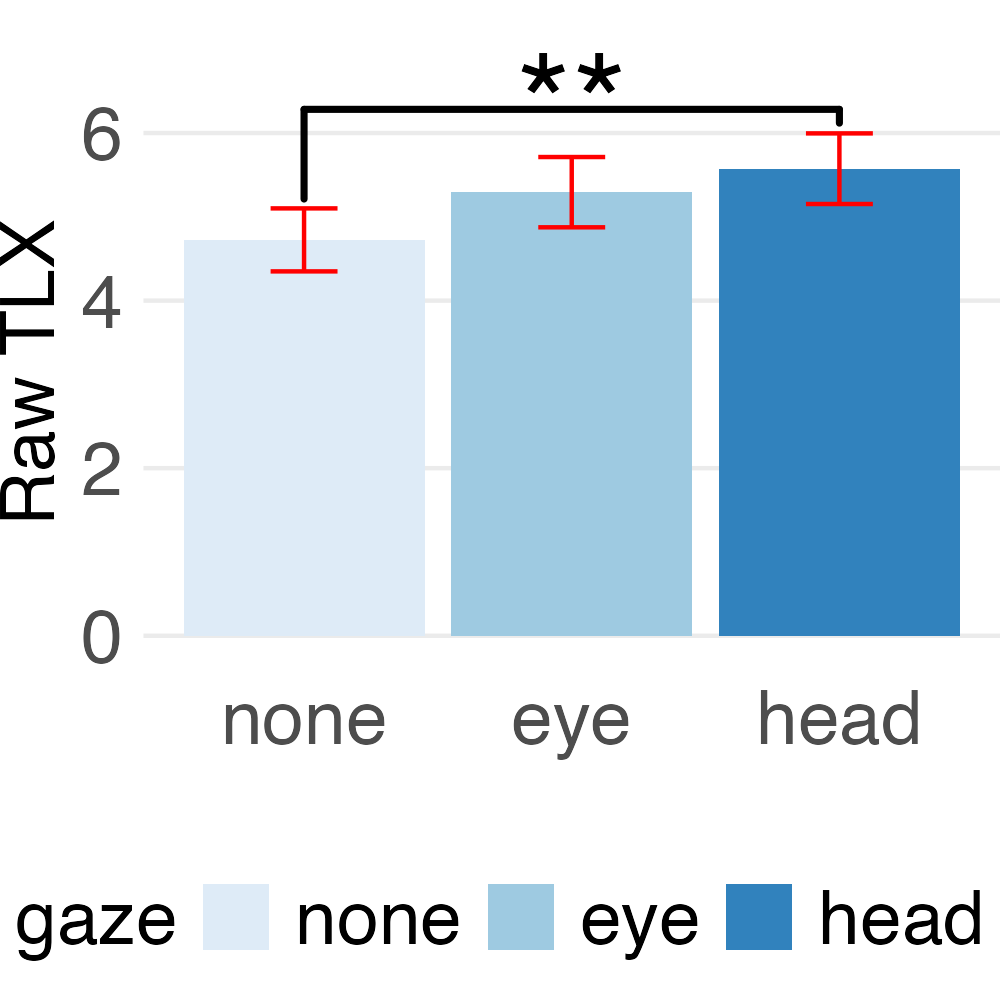}
  	\caption{Raw TLX}
  	\label{fig:results:performance:tlxG}
  \end{subfigure}%
  \\%
  \caption{Bar charts of the Raw TLX scores by \textit{perspective}~(\subref{fig:results:performance:tlxP}) and \textit{gaze}~(\subref{fig:results:performance:tlxG}). The error bars indicate the standard error.}
  \label{fig:results:performance:tlIOSSEQ}
\end{figure}

In order to measure the mental load, we used the Raw TLX using ART. The results show a significant effect on the Raw TLX for the \textit{perspective} (F$_{1,175}$ = 8.14, p$<$.01) with \textit{1PP} being less mentally demanding than \textit{3PP} (see Figure~\ref{fig:results:performance:tlxP}). Further, the \textit{gaze} also significantly affects the Raw TLX (F$_{1,175}$ = 6.73, p$<$.01). The post-hoc tests reveal a significantly higher mental load for \textit{head} than for \textit{none} \textit{gaze} (p$<$.01) as can be seen in Figure~\ref{fig:results:performance:tlxG}.

\subsection{Task Difficulty}
The task difficulty was estimated through the SEQ using ART. The analysis shows that the \textit{perspective} affects how easy the task feels (F$_{1,175}$ = 3.93, p$<$.05) with \textit{1PP} being higher rated, which means that it is perceived as easier by the participants (see Figure~\ref{fig:results:performance:IOSSEQ}).

\subsection{Social Connectedness}
\begin{figure}
  \centering
  	\includegraphics[width=\linewidth]{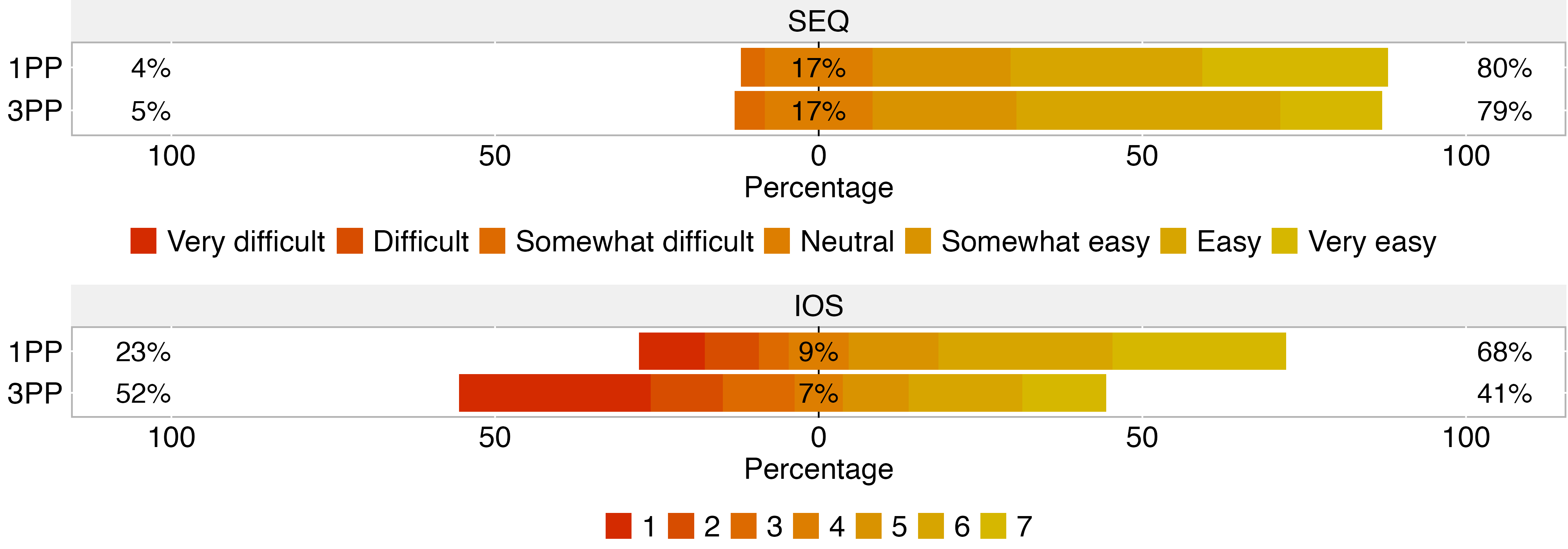}
  \caption{Likert charts of the SEQ and IOS show the percentages of responses by \textit{perspective}.}
  \label{fig:results:performance:IOSSEQ}
\end{figure}

To evaluate the users' feeling of social connectedness to the instructor, we used the IOS scale as well as the GEQ-SPM-E and GEQ-SPM-BI components. The analysis shows that the \textit{perspective} effects the IOS (F$_{1,175}$ = 52.77, p$<$.001) with the participants feeling closer to the expert in the \textit{1PP} than in the \textit{3PP} as can be seen in Figure~\ref{fig:results:performance:IOSSEQ}. For the GEQ-SPM-E ($Median = 0.83$) and GEQ-SPM-BI ($Median = 2$), we detected no significant effects using ART.

\subsection{Embodiment}
\begin{figure}[tbp]
  \centering
  \begin{subfigure}[b]{0.68\columnwidth}
  	\centering
  	\includegraphics[width=\textwidth]{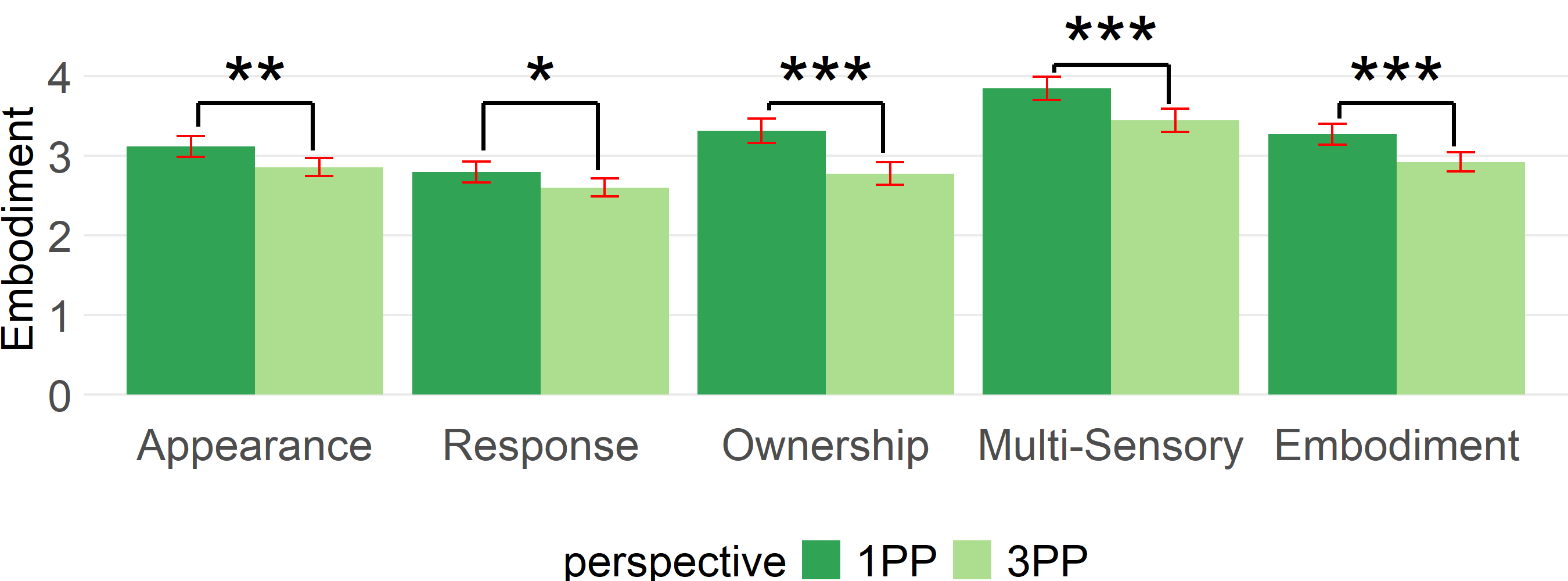}
  	\caption{Embodiment}
  	\label{fig:results:performance:embodiment}
  \end{subfigure}%
  \hfill%
  \begin{subfigure}[b]{0.32\columnwidth}
  	\centering
  	\includegraphics[width=\textwidth]{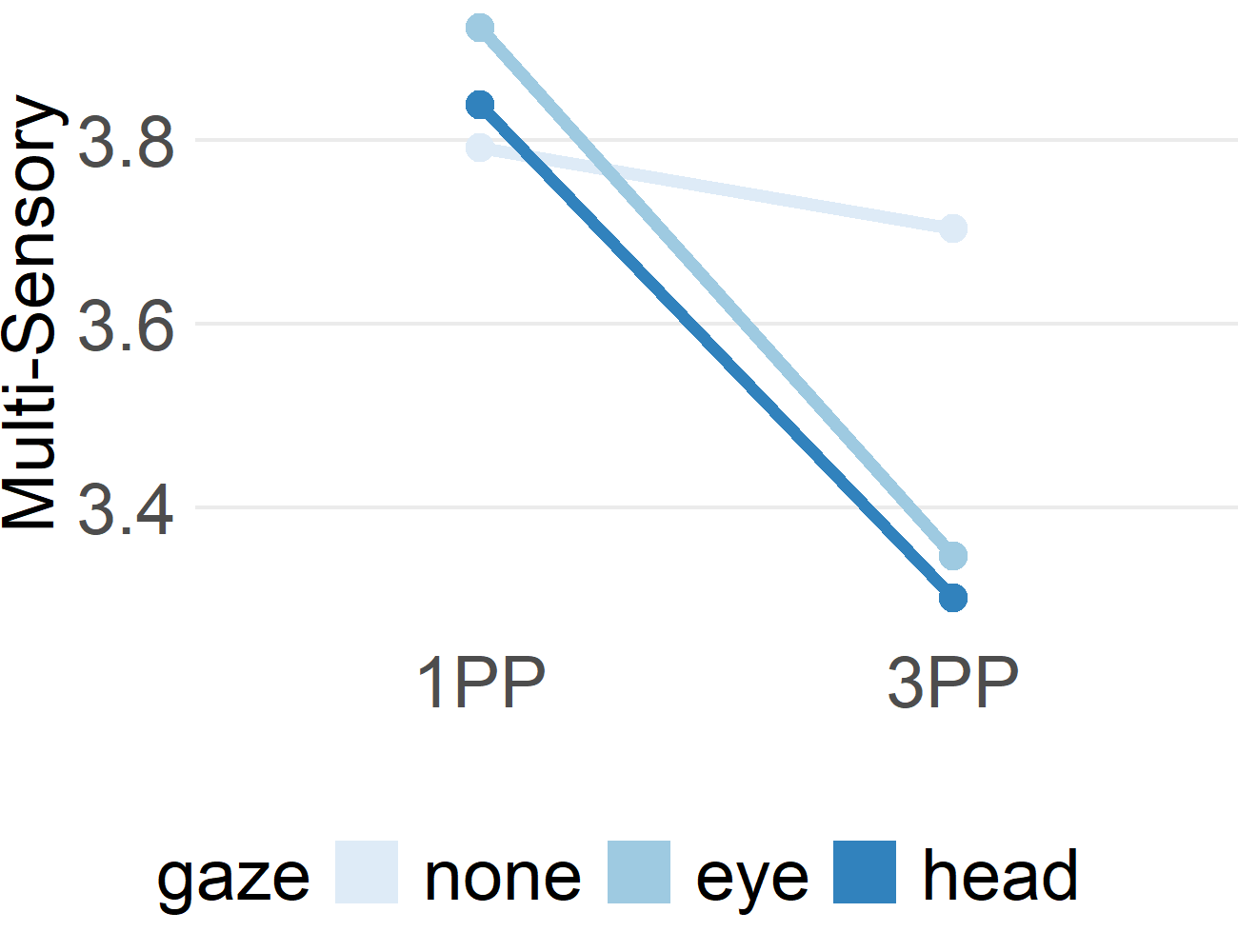}
  	\caption{Multi-Sensory}
  	\label{fig:results:performance:multi_sensory}
  \end{subfigure}%
  \\%
  \caption{Bar chart of the Embodiment subcategory scores by \textit{perspective} (\subref{fig:results:performance:embodiment}) as well as bar and interaction charts of the Multi-Sensory score by \textit{perspective}$\times$\textit{gaze} (\subref{fig:results:performance:multi_sensory}). The error bars indicate the standard error.}
  \label{fig:ex_subfigs}
\end{figure}

To assess the embodiment, we used the short version of the Avatar Embodiment questionnaire, consisting of four categories that are calculated to the overall Embodiment. The analysis indicates that the \textit{perspective} influences the Appearance (F$_{1,175}$ = 9, p$<$.01), Response (F$_{1,175}$ = 5.49, p$<$.05), Ownership (F$_{1,175}$ = 29.5, p$<$.001), Multi-Sensory (F$_{1,175}$ = 13.50, p$<$.001), and the overall Embodiment (F$_{1,175}$ = 19.66, p$<$.001) with \textit{1PP} leading to higher results (see Figure~\ref{fig:results:performance:embodiment}). In addition, we used ART for the Multi-Sensory category which shows significant effects between \textit{perspective} and \textit{gaze} for the Multi-Sensory category (F$_{1,175}$ = 3.12, p$<$.05). The post-hoc tests reveal that (\textit{1PP},~\textit{eye}) was rated higher than (\textit{3PP},~\textit{eye}) (p$<$.01) and (\textit{3PP},~\textit{head}) (p$<$.01). Also the participants feel more Multi-Sensory embodiment for (\textit{1PP},~\textit{head}) than (\textit{3PP},~\textit{eye}) (p$<$.05) or (\textit{3PP},~\textit{head}) (p$<$.05) (Figure~\ref{fig:results:performance:multi_sensory}).

\subsection{Additional Questions}
Furthermore, the participants answered questions regarding the instructions and performance after each condition.

\textbf{Instructions:}
The results show that the \textit{gaze} influences the willingness to use the interactions frequently (F$_{1,175}$ = 3.35, p$<$.05). The post-hoc test confirms a significantly higher rating for \textit{none} compared to \textit{head} \textit{gaze} (p$<$.05).
When we look at the ART analysis of the question ``\textit{I understood the instructions.}'' the \textit{perspective} has significant effects (F$_{1,175}$ = 4.23, p$<$.05) with \textit{1PP} being rated higher.
For the question "I liked the instructions" we could not observe any significant effects ($Median = 6$) using ART.

\begin{figure}
    \centering
    \includegraphics[width=\linewidth]{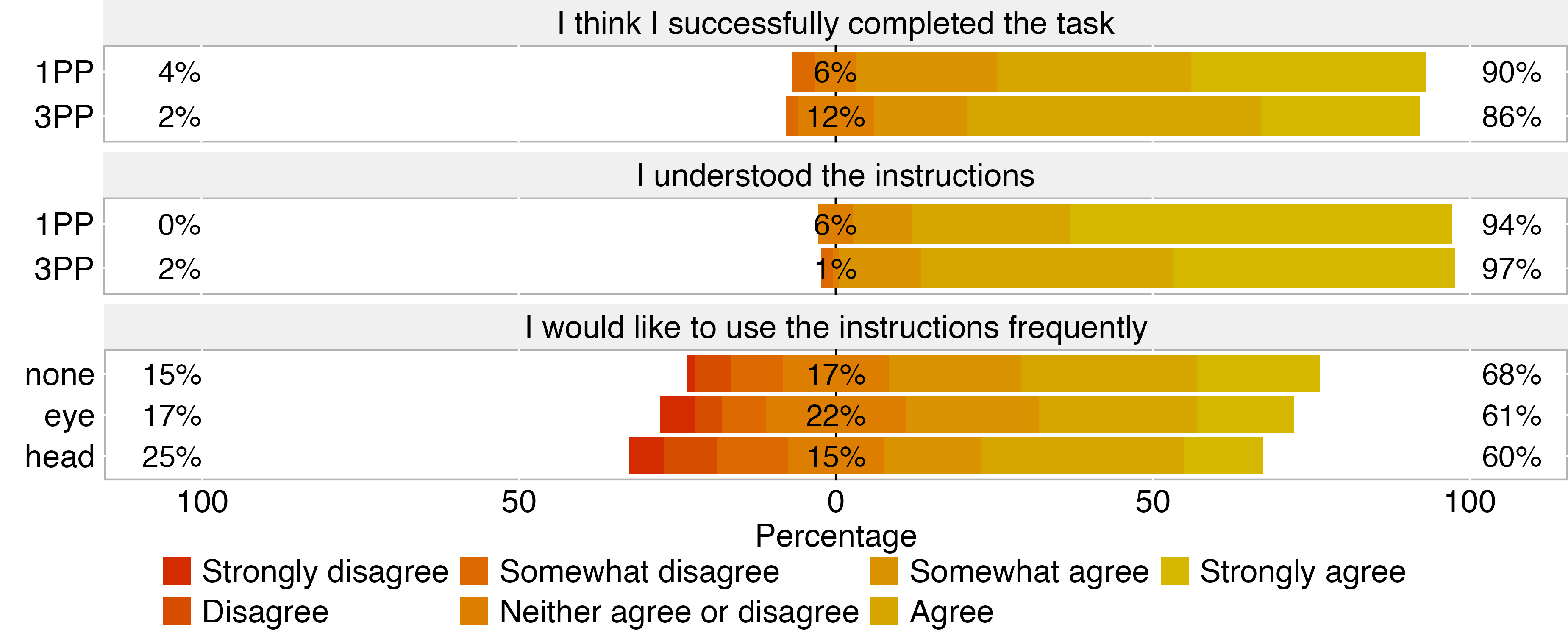}
  \caption{Likert charts of the additional questions show the percentages of responses by \textit{perspective} 
    and \textit{gaze}.}
  \label{fig:own}
\end{figure}

\textbf{Performance:}
The analysis using ART shows that \textit{perspective} has significant effects on the feeling of the participants on successfully completing the task (F$_{1,175}$ = 5.64, p$<$.05) with \textit{1PP} being graded higher than \textit{3PP}.
In addition, the ART analysis shows no effects on the question about the instructor ``\textit{I thought the other performed well.}'' ($Median = 6$). All significant differences can be seen in Figure~\ref{fig:own}.

\subsection{Interviews}
After the study, we interviewed the participants about their experiences in a semi-structured procedure while recording them. We then transcribed the audio and categorized the results. In the following, we report our findings.

For the \textbf{\textit{perspective}} that the participants \textbf{liked the most}, 69.4\% stated they prefer the \textit{1PP}, 25.0\% the \textit{3PP}, and 5.6\% were indifferent. The participants noted that the \textit{1PP} was easier (33.3\%), less distracting (13.8\%), felt more connected (2.7\%), had better graphics (2.7\%) and view (58.3\%), or that they only paid attention to the hands anyway (2.7\%). 8.3\% mentioned that the \textit{1PP} felt like a game. Others stated that the \textit{1PP} felt uncanny due to different hand features (2.7\%), that they disliked that the \textit{1PP} floating hands felt like body fragments (5.6\%), that it was hard to see the whole board in \textit{1PP} (8.3~\%), or that after experiencing the \textit{3PP} the \textit{1PP} felt like the body of the instructor was in front, in or behind their own body (16.6\%). For some, the \textit{3PP} was more familiar (22.2\%), comfortable (8.3\%), and easier to predict (2.7\%). Of the participants, 36.1\% did not like that, especially in \textit{3PP}, seeing is harder. 2.7\% perceived the \textit{1PP} and the \textit{3PP} as two separate entities.

Regarding which \textbf{\textit{gaze}} the participants \textbf{liked the most}, 30.6\% said \textit{none}, 25.0\% \textit{eye}, 19.4\% \textit{eye} and \textit{head}, 5.6\% only \textit{head}, and 19.4\% were indifferent, with 71.4\% not focusing or 28.6\% not noticing any \textit{gaze} cues. 33.3\% liked the \textit{gaze} guidance and other 30.5\% perceived the \textit{gaze} as distracting. 13.8\% did not notice a difference in \textit{gaze} size, and 5.5\% first thought the cues represented their gaze, reporting that the cursor was lagging. Of the participants (33.3\%) liked the precise guidance of the \textit{eye} and thought it was helpful and memorable (13.8\%), while 5.6\% described the \textit{eye} as stressful and distracting (2.7\%). 11.1\% liked the \textit{head}, as it enclosed both active elements and was not stressful (2.7\%). 25.0\% described the \textit{head} as not focused, confusing, and useless, as there were too many elements within the circle as well as distracting (5.6\%).

In \textbf{combination} with the \textit{1PP}, 33.3\% favored \textit{none}, 11.1\% \textit{eye}, 11.1\% \textit{head} and 16.6\% were indifferent about the \textit{gaze} cue. For \textit{3PP} 8.3\% liked the connection with \textit{none}, 5.6\% with \textit{eye}, and 5.6\% with \textit{head} the most. 2.8\% preferred the \textit{3PP} and were indifferent to the gaze. Further, 16.6\% reported they only focused on the hands and the green elements.

All participants described the \textbf{instructions} as well-made, clear, and easy to copy. 16.6\% liked the pace of the tasks, 13.8\% the length of each step, 11.1\% the haptic feeling, and 2.7\% the resemblance to real machine interfaces. Of the participants, 22.2\% reported they liked the green element highlights, with 8.3\% approving that the green elements did not immediately disappear after the animation to memorize the settings, and 8.3\% stated that they preferred the AR instructions to paper or audio instructions. Still, they disliked the fixed position during the instructions (2.7\%) and that there was no audio (2.7\%).

When asking the participants about the \textbf{instructor}, 63.8\% said that they liked the natural visual appearance.
Also, they stated that the expert made clear what to do (8.3\%) and seemed competent (5.6\%). 5.6\% perceived the hands of the instructor as good-paced, 2.7\% as easy to mimic, and 2.7\% as not disturbing. On the one hand, the participants said the hands performed realistically (8.3\%), and on the other, no realistic movements (11.1\%). Additionally, they stated that the hand size obscured the view (2.7\%) and suggested using transparent hands (2.7\%). 25.0\% of the participants said that the visual appearance of the instructor does not matter, with overall 22.2\% not paying attention to the instructor. 11.1\% reported that the avatar did not look real. 2.7\% felt annoyed by the instructor feeling like a real person but not interacting like one, and 2.7\% missed the interaction with the expert. 2.7\% felt uncomfortable as the avatar was something they could not control. 13.8\% would not change the current visual appearance of the expert, and 2.7\% recommended an avatar resembling the study conductor, so it feels more like a real person. 19.4\% did not connect with the instructor emotionally. 5.6\% of the participants did not see a resemblance to the avatar, and 5.6\% would identify more with a woman, or 2.7\% with themselves. For 5.6\%, the choice of avatar depends on the purpose, and 2.7\% would recommend different avatars for each condition to make the interaction more interesting.

For the \textbf{task difficulty}, 72.2\% of the participants described the task as easy and 27.7\% as medium difficult. The most challenging aspects included: two simultaneous actions (19.4\%), memorizing details (13.8\%), bad view (16.6\%), distracting circles (11.1\%), gauging the distance (2.7\%), and locating themselves in the room (2.7\%). Except for the plug task, the participants described as most difficult choosing the right switch (5.6\%) or button (13.8\%), setting the knob, (13.8\%), slider (2.7\%), lever (2.7\%) and the shift (8.3\%) position as well as counting the rotations of the wheel (8.3\%) and pressing the stop button (2.7\%). 2.7\% of the participants did not know at the beginning which aspects to focus on. While some felt more uncomfortable and stressed after each condition due to the weird feeling of the \textit{1PP} and time pressure (5.6\%), some felt calmer with the tasks getting easier as they got more confident over time (8.3\%).

The \textbf{general suggestions} included 94.4\% of the participants expressing positive feelings regarding the study, as they had fun and were interested, while the others were neutral. Participants suggested including feedback if they did the task right (11.1\%), adding more expressions to the avatar (5.6\%), and audio (8.3\%). 8.3\% disliked the current state of the AR pass-through, and 2.7\% the green element alignment. 2.7\% wish for a bigger field of view, and 5.6\% wish to directly follow along with the instructor. 13.8\% of the participants mentioned, without us asking, that they want to use the system to train new employees, or even themselves, for example, to learn to drive. 2.7\% liked that there is no need for a real person.

\subsection{Qualitative Notes}
In addition to the interviews, we also gave the participants the possibility to write comments at the end of each questionnaire page while taking notes during the study. Regarding the comments, the participants also mentioned the bad view either in \textit{1PP} or \textit{3PP}, as well as that the \textit{head} cue was distracting due to its size. We noticed during the study that some participants seemed surprised the first time watching the instructions, especially in \textit{3PP}. One participant even tried to greet the \textit{3PP} instructor and seemed disappointed after not receiving a response. In the \textit{3PP}, participants were careful with the HMD cable when moving positions, crouched down, or stood on their tiptoes to see the right hand of the instructor. Moreover, in the \textit{3PP}, some participants interacted with the machine interface while walking. For three participants, we had to restart the application during a condition because the HMD had AR pass-through issues by distorting the view. In these cases, we repeated the last animation step and deleted the incomplete data. In general, it seemed important to the participants to correctly complete the task.

\section{Discussion}
The results of our study provide strong evidence that the \textit{perspective} on an expert's actions in asynchronous AR influences how successful, relaxed, and self-confident we perform manual tasks. Regarding RQ1, the findings indicate that the \textit{perspective} impacts the efficiency, embodiment, and social connectedness with \textit{1PP} performing better than \textit{3PP} and is preferred by more users. About RQ2, the empirical results show that in this study setting \textit{gaze} cues can not enhance efficiency, embodiment, and social connectedness during asynchronous visual-only AR instructions, indicating that adding \textit{head} \textit{gaze} cues can improve Patience compared to \textit{eye gaze} but also mental load in contrast to using \textit{none} cue. In addition, we observed interaction effects between \textit{perspective} and \textit{gaze} (RQ3) for correct hand usage, Patience, and Multi-Sensory embodiment. In this section, we will further discuss our findings.

\subsection{Asynchronous Perspective Matters}
Related work showed that \textit{1PP}, especially in video production, has advantages in explaining practical tasks. Our work also confirms this finding for asynchronous AR instructions. The \textit{1PP} achieved in all accuracy categories the highest results as it was better understood, leading to a feeling of increased performance. A reason could be the less obscured view on the interface~\cite{thanyaditXRLIVEEnhancingAsynchronous2022}, the higher social connectedness to the instructor and embodiment~\cite{banakouVirtuallyBeingEinstein2018} compared to the \textit{3PP}. Also, in \textit{1PP}, the participants were faster when looking at the time between the beginning of the animation and the pressure of the stop button, as they had direct access to the interface, with some even beginning to interact with it before the animation ended. In the \textit{3PP}, people always need to move or turn their heads to see the instructor's actions. This does not mostly apply to the \textit{1PP} as the user and the expert are in the same position. While the individual time differences are small, they become relevant for repeated or multiple task steps. When analyzing only the time the participants interacted with the machine interface, the \textit{3PP} was faster than the \textit{1PP}, which could have influenced the accuracy, as taking less time to conduct a task can lead to more errors~\cite{garrettStudyRelationAccuracy1922a}. We told the participants at the beginning of the study to execute the instructions exactly as the expert did. Despite this, in the \textit{3PP}, some participants operated the tasks while walking, leading to less accurate imitation, more mistakes, and increased usage of only their dominant hand. This resulted in a lower task completion time with the interface as they focused less on the right order. In addition, we measured that \textit{1PP} was less mentally demanding than \textit{3PP}, which also influences the efficiency. This aligns with related work as fewer mental manipulations like spatial rotation and perspective taking~\cite{tangComparativeEffectivenessAugmented2003} are required in the \textit{1PP} than the \textit{3PP}, leading to reduced mental load. While measuring the efficiency is important, we paid particular attention to the participants' feelings during the interaction: The participants stated that the \textit{1PP} felt easier and more understandable. Still, for some, this visualization felt unnatural compared to the full-body avatar and even frightened participants to the point where they felt more pressure throughout the study due to unfamiliar floating hands. The resulting feeling of the expert standing right behind can impact the ability to remember~\cite{lordEffectsPerceivedScrutiny1987}. This raises the question for future work if other \textit{1PP} visualizations for AR would feel more pleasant while maintaining or even enhancing our measured results. As we only evaluated the \textit{1PP} during a short amount of time, we need to conduct further research on the long-term influences of using this visualization type. 
Throughout this study, we learned that taking the \textit{1PP} of an expert can improve performance in practical tasks in asynchronous XR scenarios. However, as this is a newer type of visualization of an instructor and can therefore appear unfamiliar or creepy, special emphasis should be paid to the individual preferences of the users. Future developments should therefore give the option to switch between perspectives or alter the appearance of the instructor. Further, implementations should be designed near the task without unnecessary elements obstructing the view.

\subsection{Gaze Cues can Distract}
Interestingly, because we did not tell the participants about the \textit{gaze cues} beforehand to not influence the results, many did not use or notice the \textit{gaze} visualizations of the expert. Still, the \textit{head} cue increased mental load compared to \textit{none} \textit{gaze} with the circle being too far away from the point of interest, resulting in the participants favoring the \textit{none} cue more frequently than the \textit{head}. This is contrary~\cite{bovoConeVisionBehavioural2022, piumsomboonEffectsSharingAwareness2019a} to synchronous XR studies using similar cues, indicating that the results depend on the time aspect and task. These studies used \textit{gaze} cues for collaborative analysis and visual tasks, instead of manual instructions. Therefore, choosing another task could lead to different results. More often, the participants used only one hand for \textit{head} in combination with \textit{3PP} than for \textit{none} in the \textit{1PP} condition, as the mixture of a bad view, movement, and enhanced mental load led to the forgetting of the hand sequence. As the eyes move faster and more unsteadily than the head, the expert's \textit{eye} visualization behaved accordingly, leading to impatience of the participants. Still, the interviews show that the \textit{eye} was more liked than the \textit{head} as it focused on the important areas. Also, the \textit{gaze} cues improved the Multi-Sensory feeling in the \textit{1PP} compared to the \textit{3PP} while it stayed nearly the same in the \textit{none} condition. Half of the participants preferred the \textit{gaze} cues, showing that there could be further influencing factors. 

During the study, we used the circular \textit{gaze} cues as they were suitable for both \textit{eye} and \textit{head} \textit{gaze}, while being a good intermediate between visibility and unobtrusiveness. Nevertheless, we wonder if other \textit{gaze} visualizations could be designed more suitable for this scenario. The location of the \textit{gaze} cues was on the panel to not cover elements and therefore showed no three-dimensionality, which could potentially be optimized by using, for example, beams for the \textit{eye} \textit{gaze}. Moreover, we believe that the high visual load during the tasks may be responsible for the results. Future work should therefore use smaller and less unsteady cues that still do not cover important elements. Moreover, related work showed \textit{gaze} cues often in an auditory context~\cite{piumsomboonEffectsSharingAwareness2019a, bovoConeVisionBehavioural2022}, leading to the question of whether combining them with other modalities could improve the outcome. Additionally, as the gaze behavior depends on the skill level, adjusting the cues to the user's skill could optimize performance.

\subsection{Missing Social Response}
During in-person instructions, the expert gives feedback on what is done correctly and which aspects could be improved, helping to better reflect on the performance. Even though the majority of participants perceived the avatar as realistic, had fun, liked the task and the system, and some even already wanted to use it to learn skills in the future, a downside was the instructor's inability to respond explicitly or implicitly to participants' actions. This led to disappointment and influenced the perception of the asynchronous expert. Another factor is the difference between getting advice from someone well-known, briefly known, or a stranger being presented as an avatar. Although the instructor was seen as an expert, the answers toward empathy and behavioral involvement were low. This could be a result of the participants mainly focusing on the task and a reason for not paying attention to the social \textit{gaze} cues. Also, social connectedness and efficiency could be improved by showing the instructions in parallel while doing the task instead of sequentially, which is often used in physical VR trainings~\cite{hoangOnebodyRemotePosture2016}. As a consequence, this should result in higher embodiment and accuracy by simulating physical synchronicity and reducing mental load. Implementing AI and making the asynchronous expert interact with the users visually and verbally could further change the perception and attention towards the avatar. Keeping the rapidly advancing developments in the field of generative AI in mind, future systems could interact like real instructors, giving advice and \textit{gaze} cues adapted to the users' needs while being independent of space and time.

\section{Limitations and Future Work}
Our results show that technologies offer new ways of enhancing the imitation of actions through augmentation. Still, during the study, we faced some limitations but also found new directions for future work, which we describe in the following. 

\subsection{External Validity}
Transferring user study results into real situations often comes with restrictions. Having the place fixed during the instructions is not automatically applicable in the real world, as spatial tasks are a common part of most learning processes. Compared to Cao et al.~\cite{caoExploratoryStudyAugmented2020}, we deliberately omitted this aspect in order to achieve high internal validity for two-handed manual tasks and the influence of \textit{1PP} and \textit{3PP}. Additionally, the interaction with the machine interface and the generic task itself were artificial and receptive for the purpose of the study. As successful imitation is only one of the first steps of learning~\cite{brunerFolkPedagogies1999, banduraSociallearningTheoryIdentificatory1969}, future research should therefore explore alternative conditions like conducting short-term recall of longer sequences, long-term, and field studies, to find additional influences on the learning process during asynchronous XR usage. Further, we did not track the eye movement of the participants in relation to the gaze visualizations. With the missing eye-tracking functionality and virtual element calibrating issues of the HMD, we wanted to avoid additional confusion and distraction by mounting and calibrating eye trackers before each condition. This data could give further insights about the social connectedness of the participants to the virtual instructor, at which time the participants gave attention to the different elements, and if the \textit{perspective} and \textit{gaze} cues were influencing factors.

\subsection{Diverse Visualizations}
We only included one kind of avatar in the study design to keep the conditions comparable. With people's empathy varying towards different characteristics, changing the appearance, gender, or even using scans of well-known people or the user's own body could be valuable approaches, as stated by the participants.
As already mentioned, we only used one \textit{1PP} visualization in AR to not include further conditions that could have overstretched the length of the study. We used a balanced within-subject design to keep the influences between the participants limited. Nevertheless, seeing all conditions affected the participants' perception of the expert. 

In addition, we used \textit{gaze} cues that are mainly implemented in \textit{3PP} as the \textit{1PP} and asynchronous scenarios are less explored. Therefore, more work should be invested in designing, adapting, and evaluating new \textit{gaze} visualizations for different perspectives as well as further types of communicating \textit{gaze} like audio or haptics. Moreover, we faced hardware issues with the HMD not having any marker or image tracking accessible, as well as limitations of the pass-through leading to inaccurate tracing, wobbly vision, and even system crashes. This results in the visualization feeling less real, which potentially affected the study outcome. With future technologies improving the resolution and the object tracking in the room, no further adjustments of virtual elements will be needed. Additionally, research should explore how these findings transfer to other asynchronous XR technologies, such as VR or Mixed Reality.

\subsection{View into the Future}
Creating tutorials still requires extensive and resource-consuming filming, editing, planning, an instructor's presence, and expertise in didactic 3D knowledge transfer. These tutorials often lack adaptability to user behavior and can miss important learning factors. Valuable social cues can get lost or overwhelming due to the wrong representation for the individual learner. As technology rapidly advances, we expect that capturing, displaying, and adapting three-dimensional information can be largely automated in the near future, leading to diverse applications in the direction of our work.

Current scenarios focus on recording interactions of a skilled person for replay when explanations or memory aids are needed. This process is time-intensive, requiring re-filming for every new task or machine update. Generative AI can change this by enabling instant tutorial creation from descriptions, device visuals, or interaction recordings, generalizing usage scenarios across interfaces. Social cues can be preserved and adapted to the needs of individual users, avoiding manual effort. Given such a system, the virtual expert can predict~\cite{zhengEgocentricEarlyAction2023} and properly respond to the user's actions. However, this outlook raises several new questions, such as: What are the important aspects that need to be maintained, and which information is necessary? How do the users and their surroundings feel when getting part of the tutorial of others? What are the consequences when such tutorials are used in an unethical way? Future research should also ensure the privacy and security of the expert's information, as well as define guidelines and visualization techniques so that these factors are not violated. Nevertheless, we emphasize that our work continues to be valid for these visionary scenarios, and we expect the influence of perspective and gaze to remain relevant.
Therefore, we are confident that our work serves as a foundation for future approaches to solving these questions.

\section{Conclusion}
In this paper, we explored the influence of asynchronous XR expert \textit{perspective} and \textit{gaze} visualizations on efficiency, embodiment, and social connectedness while users operated a manual task. In an experimental setup, we tested in a quantitative and qualitative within-subject user study 36 participants, and collected empirical data. To conduct the experiment, we built a machine interface with physical controls and an AR application of a specialist showing two-handed tasks that the user had to imitate after each animation. Our findings suggest that the \textit{1PP} improves efficiency when replicating the task as well as embodiment and social connectedness towards the virtual expert. Further, we could show that \textit{gaze} cues do not improve the measured variables during visual-only demonstrations in our setting. Still, half of the participants stated that they preferred the \textit{gaze} cues. We recommend future research in this area to explore further use cases, \textit{perspective}, and \textit{gaze} visualizations while also investigating AI and privacy considerations in the field of asynchronous XR instructions. Our work represents a great step in the direction of effective three-dimensional knowledge transfer that not only includes social factors but is also enjoyable for the users.

\acknowledgments{
This work was funded by the Deutsche Forschungsgemeinschaft (DFG, German Research Foundation), Project-ID 517713394 and the LOEWE initiative (Hesse, Germany) within the emergenCITY center [LOEWE/1/12/519/03/05.001(0016)/72].}

\bibliographystyle{abbrv}

\bibliography{Asynchronous_Assistence}
\end{document}